# Stretching and Compressing Capillary Bridges on Hydrophilic, Hydrophobic, and Liquid-infused Surfaces

Sarah Jane Goodband[1,⊥], Ke Sun[1,⊥], Kislon Voïtchovsky[1*], Halim Kusumaatmaja[2*]

[1]Department of Physics, Durham University, Durham, DH1 3LE, UK

[2]Institute for Multiscale Thermofluids, School of Engineering, The University of Edinburgh, Edinburgh, EH9 3FB, UK

,⊥ These authors contribute equally
* Correspondence: kislon.voitchovsky@durham.ac.uk, halim.kusumaatmaja@ed.ac.uk

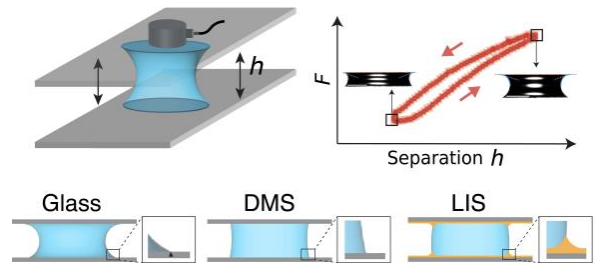

**Abstract**: Aqueous capillary liquid bridges are ubiquitous in nature and in technological processes. Here, we comparatively investigate capillary bridges formed between three distinct types of surfaces: (i) hydrophilic glass, (ii) hydrophobic dichlorodimethylsilane (DMS)-functionalized glass, and (iii) silicone-oil-infused LIS. We combine experimental measurements and computer simulations of the capillary bridges' evolution upon changes in the gap size between the surfaces, deriving in each case the bridge geometry and the resulting capillary force. The results, also compared with predictions from the existing theory, follow expected trends on glass and DMS-functionalized surfaces: contact line pinning dominates the bridge behavior on glass with a characteristic stick–slip motion, whereas a pronounced advancing and receding hysteresis is observed on DMS surfaces. On LIS, the absence of pinning leads to minimal force variation, gravity-driven breaking of the bridge symmetry, and possible liquid exchange between LIS through bridge cloaking. These effects become particularly significant in asymmetric bridge configurations combining LIS and DMS surfaces, where the transfer of lubricant from LIS to DMS modifies the effective surface tension and alters bridge–surface interactions. Our systematic comparison of the capillary bridge behavior across solid and liquid interfaces with varying wettability provides a foundation for designing functional surface applications with controlled bridge–surface interactions.

# Introduction

Capillary liquid bridges form when a liquid meniscus bridges two surfaces, generating strong adhesion forces. They are ubiquitous in nature and industry, from insect adhesion to water surfaces,[1,2] to cohesion in soil and granular media such as sandcastles,[3,4] semiconductor fabrication,[5] oil recovery,[6] cement drying,[7] and drug delivery[8,9]. At the microscale, these capillary forces can be dominant, also influencing macroscopic properties and temporal evolution of the system.[10–12] The behavior of capillary bridges is governed by the liquid's properties, bridge dimensions, the chemical and topographical characteristics of the surfaces, and environmental conditions such as temperature and humidity. These effects are reflected in extensive studies examining the impact of surface geometry,[4,13] chemical and topographical patterning,[14,15] wettability,[16,17] and length scales from nanometers[18–20] to millimeters[21,22]. To uncover the underlying physics, studies usually quantify the bridges geometrical characteristics (e.g. contact angles (CAs), curvatures, and contact radius)[15,23,24] and measure the capillary forces they exert[22,25,26] when they are extended or compressed between solid surfaces.

Despite substantial progress, our understanding of capillary bridge behavior remains incomplete. Most previous studies have focused on hydrophilic solid surfaces, where contact line pinning dominates and induces hysteresis.[23,24,27] More recently, investigations on hydrophobic surfaces with liquid features are emerging, reflecting their growing potential in applications such as self-cleaning and anti-fouling technologies. However, most studies hitherto employ solid surfaces, and capillary bridges between liquid-like surfaces are still largely unexplored. Liquid-infused surfaces (LIS) present a typical example of such surfaces, where porous structures are impregnated with a lubricant to achieve high liquid repellency and low friction properties.[28,29] From a fundamental perspective, LIS represent a distinct capillary behavior regime where a three-phase interface is present. From the application side, advances in LIS and related liquid-like functional surfaces such as slippery omniphobic covalently attached liquid (SOCAL) surfaces, are critical in liquid deposition and transport in fields such as anti-icing coatings, inkjet printing, and microfluidics.[30–32] On LIS, pinning and hysteresis are negligible,[33,34] and the introduction of a lubricant creates more liquid–liquid and liquid–gas interfaces. This can alter capillary morphology and cause deviations from classical force models. Computational modelling by Shek et al. has shown lubricants on LIS can produce fundamentally different capillary geometries compared to solid surfaces, with increased vertical friction arising from oil ridges formed around the contact of the bridge and the surface.[35] Furthermore, the fluid nature of LIS could induce other unexplored phenomena, such as lubricant transport between surfaces via capillary bridges.

Here, we quantitatively compare capillary bridges evolution during extension and compression on hydrophilic, hydrophobic, and LIS using a micronewton-precision experimental setup. Experiments are complemented by computational modelling that incorporates apparent contact angle to account for oil ridge formation on LIS, which is experimentally challenging to capture but crucial for influencing bridge geometry and forces. Silicon oxide glass is selected as the hydrophilic surface for its routine use and technological relevance, while DMS-functionalized hydrophobic surface and silicone-oil-infused LIS are selected for their similar wettability to isolate effects specific to the fluid nature of LIS. Comparison across these three surface types presents distinct force and geometry responses driven by phenomena such as stick–slip motion, contact angle hysteresis, and liquid ridge formation. Tests on dissimilar surface pairs of DMS and LIS further demonstrate bridge asymmetry induced by gravity under small capillary forces, as well as lubricant transfer from LIS to the opposing surface. Overall, we present a systematic experimental and computational study that establishes a benchmark for understanding and predicting capillary bridge evolution on solid and liquid functional surfaces, offering mechanistic insights to the rational design of surfaces with liquid features.

## Methods

**Surface Preparation**

**Hydrophilic surface – Glass.** Silicon oxide glass coverslips (25 mm × 25 mm, thickness 0.13–0.16 mm, VWR, UK) are used directly from a freshly opened box without additional cleaning procedures to ensure chemical stability during measurements. We characterized the surface roughness using atomic force microscopy (AFM), obtaining a mean surface roughness $S_a = 0.449 \pm 0.027$ nm, a root-mean-square roughness of $S_q = 1.031 \pm 0.207$ nm, and a surface area $A$ to projected area $A_0$ ratio of $A/A_0 = 1.00039 \pm 0.00008$ (see details in Supporting Information SI 1).

**Hydrophobic surface – DMS.** Hydrophobic surfaces are prepared by chemical vapor deposition (CVD) of dichlorodimethylsilane (DMS) on glass coverslips (reference hydrophilic surfaces).[36] Glass slides are sequentially cleaned by acetone (99%, Sigma-Aldrich, UK) and isopropanol (99.8%, Fisher Scientific, UK), followed by 30 min of sonication. They are then dried under nitrogen and plasma-cleaned for 10 min (>30 W, VacuLAB-X, UK) and are dehydrated in an oven at 100 °C for 1 h. For CVD, 1 mL of DMS is placed in an open dish in a desiccator along with the slides directly transferred from the oven and kept under vacuum overnight. Finally, the slides are rinsed with acetone and ultrapure water (18.2 MΩ, Merck-Millipore, UK), then dried at 40 °C overnight. AFM measurements of the DMS-functionalized surface yield $S_a = 2.297 \pm 0.221$ nm, $S_q = 3.427 \pm 0.380$ nm, and $A/A_0 = 1.00037 \pm 0.00007$ (see details in Supporting Information SI 1).

**LIS.** LIS are prepared following established protocols.[37,38] In short, glass slides are firstly cleaned by soaking in an aqueous solution of Decon 90 (Decon Laboratories Ltd., UK) before rinsing and sonicating in ultrapure water to remove residual detergent, followed by air-drying. Prior to coating, slides are rinsed with acetone and isopropanol and dried under nitrogen and then air-dried. Five layers of nanoparticles are then sequentially applied to the surface using a liquid spray (GLACO™, SOFT 99 Corp.) with 1 h interval between layers. A drop of 50 µL silicone oil (20 cSt @ 25 °C, density $\rho = 0.95\ g\ mL^{-1}$, surface tension $\gamma = 20.6\ mN\ m^{-1}$ in air, Sigma-Aldrich, UK) is then placed on the surface and spin-coated (2000 rpm, 5 min). The slides are used immediately or stored without oil coating, in closed Petri dishes for a maximum of 2 weeks. LIS fabricated with this protocol retains a stable oil layer thickness (>3 µm) with no exposed nanoparticles. This allows the LIS to maintain their chemical and wetting properties over the timescale of the experiments,[38] as also evidenced by the negligible hysteresis reported in the Results section.

**Capillary bridge measurement and error control**

A detailed description of the experimental protocol, apparatus, and data processing procedure is provided in Goodband et al.[39] In brief, the capillary bridge is formed between two parallel substrates, with the top plate attached to a force sensor and the bottom plate mounted on a motorized stage used to compress and stretch the bridge. The system is imaged using a dual camera arrangement: one camera focuses on the bridge edge to provide high resolution profiles for extracting geometrical parameters, while a second synchronized camera views the bridge from another side to support the analysis (see an illustration in Supporting Information SI 2). We maintain identical capillary bridge volume and composition across all experiments to ensure direct comparison.

Experiments are conducted under ambient laboratory conditions (20–23 °C, relative humidity 60–70%). To start with, two solid surfaces prepared using the above protocols are mounted onto a custom-built plate using an adhesive (Reprorubber™, Bowers Group, UK) and are allowed to cure for 2 h before the plate is mounted to the force sensor. The force sensor is then equilibrated for 1 h before measurements. To ensure protocol consistency and data reproducibility, a 10 μL droplet of 80 wt% glycerol in ultrapure water (used to limit evaporation) is placed always onto the bottom surface and is gently brought into contact with the top surface (see details in Supporting Information SI 3), followed by equilibration for 2 min. The droplet has a measured density $\rho = 1.20\ g\ mL^{-1}$, a viscosity $\eta = 45.35\ mPa\ s$ (from Moreno-Labella et al.[40]), and a measured surface tension $\gamma$ of $67.2\ mN\ m^{-1}$ in air and $27.6\ mN\ m^{-1}$ in 20 $cSt$ silicone oil. During measurements, the bridge is firstly extended and then compressed at a constant rate of $0.008\ mm\ s^{-1}$, with a maximum separation difference of $0.5\ mm$ between the most compressed and most stretched states."

Accurate tracking of the capillary bridge contact line is important for understanding pinning effects but is experimentally challenging to achieve simultaneously at both the top and bottom interfaces due to limitations in optical focus (see Supporting Information SI 4). However, capturing information at both extremities is necessary for quantifying the gravitational effects, probing asymmetric bridges, and evaluating the experimental approach against theoretical predictions. In practice, we acquire top and bottom measurements separately by readjusting camera to capture top and bottom in subsequent extension-compression cycles, with control experiments confirming that the measured values do not vary significantly (see Supporting Information SI 5).

Even following the same protocol, variations in the measured geometrical parameters can be observed between data sets. For instance, changes in ambient humidity and temperature can result in CA variations of ~2° for a LIS sample[38], and DMS surfaces are observed to have up to 3° CA differences following the same protocol. In all cases, the measurements are conducted over a few hours to minimize environmental impact and ensure highly consistent data sets.

## Capillary force calculation

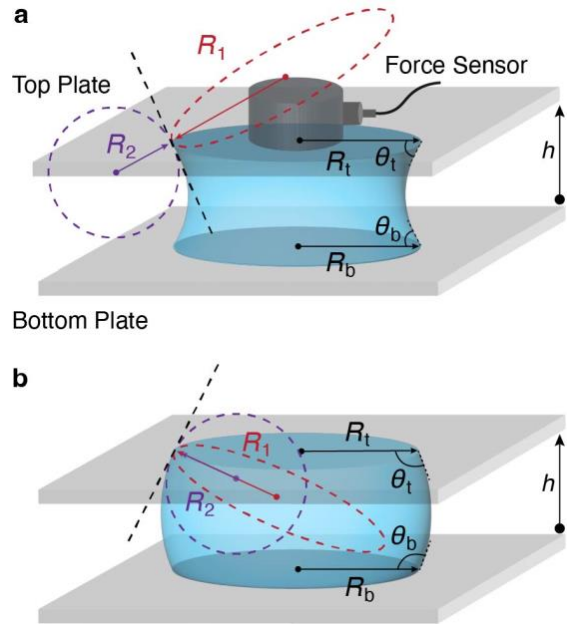

**Figure 1.** Schematics of concave (a) and convex (b) capillary bridges between two parallel plates separated by a distance $h$. $\theta_t$ and $\theta_b$ denote the contact angles at the top and bottom of the bridge, $R_t$ and $R_b$ are the corresponding top and bottom contact radii. The mean curvature of the bridge is determined from the azimuthal ($R_1$) and meridional ($R_2$) radii of curvature. $R_1$ and $R_2$ are obtained orthogonally at either the top (illustrated here) or the bottom of the bridge, depending on which plate the force is being calculated for (see details in the force derivation and Supporting Information SI 6).

To quantify the capillary force exerted by a capillary bridge, it is necessary to measure its geometry throughout the experiment. Fig. 1 shows the key geometrical parameters of a capillary bridge formed between two parallel substrates. Forces are measured exclusively on the top substrate, but they can theoretically be calculated for both.

When gravity is negligible, e.g. for a bridge much smaller than the capillary length, the equilibrium capillary force $F$ between two identical parallel plates can be expressed by the direct action of surface tension and Laplace pressure:[11,35]

$$F = -2\pi\gamma R \sin(\theta) + \pi R^2 \Delta P, \qquad (1)$$

where $\gamma$ is the liquid surface tension, $R$ is contact radius, $\theta$ is the contact angle, and $\Delta P$ is the Laplace pressure between the bridge and surrounding fluid. When gravity is negligible, the top and bottom contact angles and contact radii are equal. When gravity cannot be neglected, the capillary bridge becomes asymmetric, and forces exerted by the top and bottom differ. Following the Young-Laplace equation, the capillary force on the top plate can be calculated from the geometrical parameters at the top:

$$F_t^{calc} = -2\pi\gamma R_t \sin(\theta_t) + \pi R_t^2 \gamma \left(\frac{1}{R_1} + \frac{1}{R_2}\right) \quad (2)$$

where $R_t$ and $\theta_t$ are the top contact radius and contact angle. The radii of curvature $R_1$ and $R_2$ correspond to the azimuthal and meridional radius of curvature, respectively. For a given plate (top or bottom), the two corresponding radii are measured orthogonally at that plate: $R_2$ is obtained by fitting the local bridge edge to a second order polynomial, and $R_1$ is determined from the geometry of the three-phase contact line (see details in Supporting Information SI 6).[39] For a concave capillary bridge, $R_1$ is positive and $R_2$ is negative; for a convex capillary bridge, both $R_1$ and $R_2$ are positive. Hereafter, this equation will be called the 'top calculated force' $F_t^{calc}$. Similarly, the bottom capillary force can be expressed as:

$$F_b^{calc} = -2\pi\gamma R_b \sin(\theta_b) + \pi R_b^2 \gamma \left(\frac{1}{R_1} + \frac{1}{R_2}\right) \quad (3)$$

where $R_b$ and $\theta_b$ are the bottom contact radius and contact angle. Since this expression relies on the bottom measured parameters, it will hereafter be called the 'bottom calculated force' $F_b^{calc}$. In Eqs. 2–3, $\theta_t$ and $\theta_b$ denote the Young's contact angles on solid substrates, or the apparent contact angles in the presence of oil ridges, as defined in Section 2.4.

In the experimental setup, a force sensor is implemented on the top plate to acquire a 'top measured force', $F_t^{meas}$. While the bottom force cannot be measured directly, it can be easily inferred from the top measured force by accounting for gravity:

$$F_b^{inf} = F_t^{meas} + \rho g V \quad (4)$$

where $\rho$ is the density of the droplet, $V$ is the capillary bridge volume, and $g$ is the gravitational acceleration. Since this approach relies on the top measured force to infer the bottom force, it will be referred to as the 'bottom inferred force' $F_b^{inf}$. Eq. 3 and Eq. 4 thus provide two complementary methods to determine the bottom force, the advantages of which are discussed in detail in the Results section.

**Computational model**

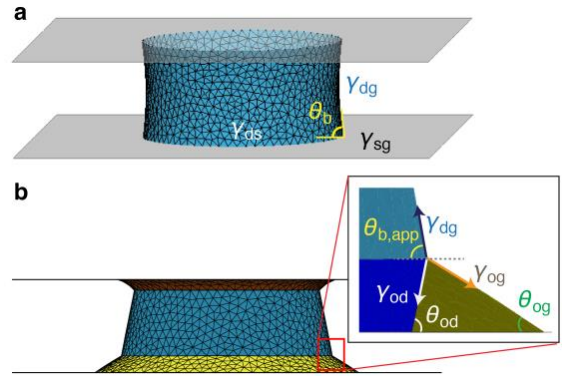

**Figure 2.** Simulation snapshots of capillary bridges between two solid parallel plates (a) and between plates with oil rings (b). $\gamma_{ds}$, $\gamma_{dg}$, $\gamma_{sg}$, $\gamma_{od}$, and $\gamma_{og}$ are the interfacial tensions of the droplet–solid, droplet–gas, solid–gas, oil–droplet, and oil–gas interfaces, respectively. $\theta_b$ in (a) is the Young's contact angle of the capillary bridge on the bottom solid plate, and $\theta_{b,app}$ in (b) is the apparent contact angle of the bridge on the bottom plate when surrounded by an oil ridge. The inset in (b) illustrates the Neumann triangle at the droplet–air–oil contact, with $\gamma_{dg}$, $\gamma_{og}$, and $\gamma_{od}$ representing the interfacial tensions that satisfy the force balance. At the bottom substrate, $\theta_{og}$ denotes the oil–air contact angle at the plate, and $\theta_{od}$ denotes the oil–droplet contact angle at the plate. The schematics in (a) and (b) explicitly illustrate the bottom plate; the same conventions apply to the top plate.

We employ quasistatic simulation using the Surface Evolver[41] software. In brief, the model incorporates the fluid and solid interfaces, with vertices relax in a gradient descent manner to reach the system's minimum energy configuration. As the capillary bridges considered here are comparable in size to the bridge liquid's capillary length, gravity is incorporated into the model by matching the Bond number to the experimental value (see details in Supporting Information SI 7). For simple solid surfaces (Glass and DMS), we initialize the droplets in between two plates, using experimentally measured droplet–gas interfacial tension $\gamma_{dg}$ and the bottom Young's contact angle $\theta_b$ (Fig. 2a). The rest of interfacial tensions are related via Young's equation, $\gamma_{ds} = \gamma_{sg} - \gamma_{dg} \cos \theta_b$, where $\gamma_{ds}$, $\gamma_{sg}$ are the droplet–solid and solid–gas interfacial energies, respectively. For cases involving LIS or lubricant transfer from LIS to DMS, the Young's contact angle on a solid surface is no longer applicable. Instead, we define an apparent droplet contact angle, such as $\theta_{b,app}$ in Fig. 2b, obtained by measuring the angle between the droplet–air interface profile and the horizontal plane. Here, we take the bottom plate as an example; a similar approach can be applied to the top surface when an oil ridge is present. In such situations, oil ridges form around the capillary bridge at its contact with the plate, giving rise to a Neumann triangle at the oil–droplet–air three-phase interface (Fig. 2b, inset). Although these features are too small to resolve experimentally, simulations that incorporate the relevant interfacial tensions can infer the three phase contact geometry. Specifically, the interfacial tensions $\gamma_{od}$ (oil–droplet), $\gamma_{dg}$ (droplet–gas), $\gamma_{og}$ (oil–gas) are obtained from pendant drop measurements. The oil–gas contact angles $\theta_{og}$ is either assumed from the intrinsic oil wettability on the substrate or independently determined from lubricant cloaking measurements (Supporting Information SI 8). These quantities are related through the following expression:[35]

$$\cos\theta_{b,app} = -\cos\theta_{od}\frac{\gamma_{od}}{\gamma_{dg}} + \cos\theta_{og}\frac{\gamma_{og}}{\gamma_{dg}} \qquad (5)$$

Eq. 5 allows $\theta_{od}$ to be derived once $\theta_{b,app}$, $\theta_{og}$, and the interfacial tensions are known. The resulting values of $\theta_{od}$, together with the measured interfacial tensions and a prescribed oil ridge volume, are then used as inputs for the simulation model. For LIS-LIS systems, $\theta_{od}$ remains essentially constant with only minor variation (~2–3°) during compression and extension. In DMS-LIS systems, the LIS side again shows little change, whereas the DMS side varies strongly, by ~30° between the most compressed and extended configurations due to lubricant transfer and ridge pinning. The actual size of the oil ridges depends on factors such as lubricant pressure and oil thickness, which could not be fully captured in the current experimental setup. To satisfy simulation resolution and system symmetry, the oil ridges input in the model may therefore be larger than those in the experiments. Nevertheless, this approximation is based on experimental measurements and yields good agreement between simulated and experimentally measured capillary bridges, indicating that both the oil ridges and the three-phase contact are accurately captured by the model (see Results section).

The simulations also allow calculation of the exerted capillary force. For convenience, here we evaluate it at the bottom contact line between the droplet and the oil ridge. Adapted from Eq. 1, the force is given by:

$$F_{sim} = -2\pi\gamma_{dg}R_{cl}\sin(\theta_{b,app}) + \pi R_{cl}^2\Delta P, \qquad (6)$$

where $R_{cl}$ is the contact line radius between the bridge and the oil ridge, and $\Delta P$ is the pressure difference across the droplet body, obtained directly from the simulation output. This formulation links the simulated interfacial geometry to the measurable capillary force, and the resulting predictions are in good agreement with experimental measurements (see Results section). Having clarified how the macroscopic contact angle is measured at for the bridge at both solid surfaces and surfaces with oil ridges, we hereafter simply refer to the angle as $\theta_t$ and $\theta_b$, regardless of whether it corresponds to the Young's contact angle on a solid substrate or the apparent contact angle in the presence of an oil ridge.

## Results and Discussions

**Capillary bridge between identical parallel surfaces**

To build a basic understanding of the capillary behavior on distinct surfaces, we begin by comparing capillary bridges between identical top and bottom surfaces, ranging from the hydrophilic Glass surface, to the hydrophobic DMS-functionalized surface, and the LIS. For clarity, hydrophilic Glass surface and DMS-functionalized surface will hereafter be referred to as 'Glass' and 'DMS' respectively. Upon extension and compression of the capillary bridges, we simultaneously measure and calculate geometrical parameters and exerted forces. Figs. 3a–c present the most compressed and most stretched geometries of the capillary bridge, showing good agreement between experiments and simulations for all cases. The bridge geometry differs across surface types, exhibiting convex or concave shapes during extension and compression. To quantify these geometrical variations, we track the evolution of the contact angles, contact radii, and the meridional curvature as a function of plate separation.

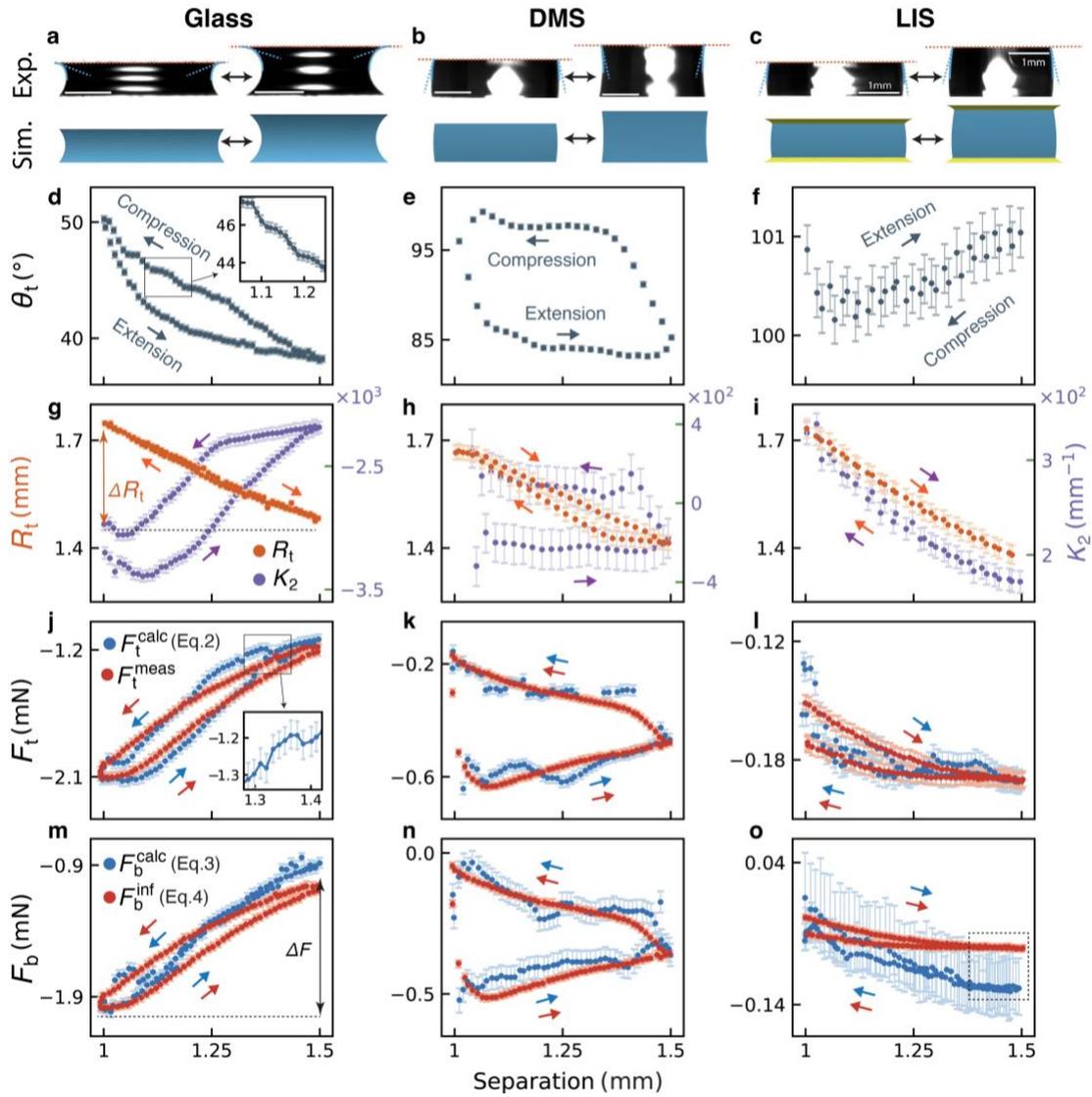

**Figure 3.** Geometry and force comparison of capillary bridges between identical top and bottom surfaces: hydrophilic glass surface, hydrophobic DMS-functionalized surface, and Liquid-infused surfaces (LIS). (a–c) Capillary bridge geometries in the most compressed and most stretched state. Evolution of the geometrical features is shown for top contact angles $\theta_t$ (d–f), top contact radius $R_t$, and meridional curvature $K_2$ (g–i, $K_2 = 1/R_2$ and $R_2$ is the meridional radius of curvature in Fig. 1). The change in contact radius within an extension-compression cycle, $\Delta R_t$, is marked in (g) as an example. Panels (j–l) compare the measured ($F_t^{meas}$, red) and calculated force ($F_t^{calc}$, blue) for the top surface, while panels (m–o) compare the inferred ($F_b^{inf}$, red) and calculated force ($F_b^{calc}$, blue) for the bottom surface. Arrows indicate the extension-compression direction in each panel, with $\Delta F$ denoting the force variation in the process, as shown in (m). Insets in (d) and (j) highlight the stepwise stick–slip features in the measured contact angles and forces. The dashed square in (o) marks the deviation between the inferred and the calculated bottom forces, particularly at larger plate separations. Error bars represent standard errors.

We first discuss the top contact angle $\theta_t$ and top contact radius $R_t$. On Glass, $\theta_t$ decreases as the plate separation increases, spanning a hydrophilic range of $38 - 50°$ in conjunction with contact line motion (Figs. 3d, 3g). Hysteresis is observed between extension and compression cycles. Notably, a stepwise increase in $\theta_t$ occurs during capillary bridge compression (Fig. 3d, inset), without dominant plateaus as the contact line advances or recedes. However, this stepwise feature is not observed in the contact radius $R_t$ (Fig. 3g). This observation indicates a complicated stick–slip behavior involving alternating pinning and rapid movements of the contact line, likely due to small and asymmetric local pinning points as the evolutions of contact angle and radius are not straightforwardly correlated. The effect of pinning is particularly evident when comparing the change in contact radius $\Delta R_t$. Glass exhibits the smallest contact radius variation ($\Delta R_t = 0.27$ mm, Fig. 3g), whereas LIS, which exhibits no pinning, shows the largest $\Delta R_t = 0.4$ mm (Fig. 3i). For DMS, the CA exhibits typical hysteresis expected for hydrophobic surfaces[23,27,42,43], with a hydrophobic (> 90°) advancing CA at ~98° and a hydrophilic (< 90°) receding CA at ~84°, corresponding to the plateaus in CA observed during compression and extension, respectively (Fig. 3e). In contrast, LIS benefits from its low friction liquid characteristics, yielding a highly stable $\theta_t$ with negligible hysteresis within experimental error (Fig. 3f). This observation aligns with other studies[44,45] using silicone oil as the lubricant, which also report low contact angle hysteresis on LIS. During extension, the slight increase in $\theta_t$ at larger separation distance (Fig. 3f) can be attributed to interactions between the capillary bridge and the LIS lubricant ridge. This is addressed further in Fig. 5 and associated text.

The meridional curvature $K_2$ ($K_2 = 1/R_2$, where $R_2$ is the meridional radius of curvature in Fig. 1) is closely related to the CA and the overall capillary bridge geometry. Among the three cases, the capillary bridge on Glass experiences the highest curvature at $\sim -3000$ mm$^{-1}$ on average (Fig. 3g), due to its low CA and significant contact line pinning. The curvature remains negative, thus the capillary bridge on Glass retains a concave shape during extension and compression. On DMS, $\theta_t$ crosses between the hydrophilic and hydrophobic regimes at 90°, resulting in both positive and negative curvatures. The capillary bridge is concave when most stretched and convex when most compressed (Figs. 3b, 3h). As a hydrophobic surface with similar CA, the meridional curvature of the capillary bridge on LIS is of similar magnitude to that on DMS but remains positive, reflecting the relatively constant $\theta_t$ and negligible hysteresis (Fig. 3i). Overall, curvature hysteresis is highest on Glass, intermediate on DMS, and negligible on LIS, consistent with the observed contact angle hysteresis.

Aside from the geometrical parameters presented in Figs. 3d–i, it is useful to consider the symmetry of the contact line during an extension-compression cycle, as capillary bridges do not necessarily move symmetrically when pinning occurs. This can be quantified by tracking the displacement of the contact points at the top and bottom of the capillary bridge with the solid plates after full extension-compression cycles, thereby allowing quantification of asymmetry on both sides. Only small displacements are observed for Glass and LIS due to strong pinning in the former and frictionless motion in the latter. In contrast, DMS exhibits a much larger asymmetry, resulting from a combination of pinning and contact line displacement (see details in Supporting Information SI 9). These observations are consistent with the expected behavior for each system, highlighting the various phenomena at play in capillary bridge behavior.

From the geometrical parameter evolution during compression and extension, we can now calculate the associated capillary force exerted by the substrates using Eqs. 2–4. The forces calculated from geometrical measurements are denoted as $F_t^{calc}$ (Eq. 2) and $F_b^{calc}$ (Eq. 3) for top and bottom surfaces, respectively. The force directly measured by the force sensor on the top surface is denoted as $F_t^{meas}$, while the bottom force is inferred by adding a gravity term, and denoted as $F_b^{inf}$ (Eq. 4). Figs. 3j–o compare measured or inferred forces with the calculated values for each system. Generally, the top measured force $F_t^{meas}$ agrees well with the calculated force $F_t^{calc}$ within experimental error, including on LIS (Figs. 3j–l). This suggests that capillary theory developed for solid surfaces can be readily adapted to predict the capillary forces on LIS, at least in the limit of a small lubricant ridge considered in this work.

The magnitude of the capillary force, however, varies significantly across these different systems. On Glass, the stronger interactions between the capillary bridge and the surface yield an absolute force value of $F \sim 2$ mN with a variation of $\Delta F \sim 1$ mN over an extension-compression cycle (Figs. 3j, 3m). In contrast, the force magnitudes on the DMS and LIS are considerably smaller, on the order of 0.1 mN. Notably, the force variation over an extension-compression cycle is five times larger for DMS ($\Delta F \sim 0.4$ mN, Figs. 3k, 3n) than for LIS ($\Delta F \sim 0.06$ mN, Figs. 3l, 3o), thanks to the frictionless nature of LIS. On Glass and DMS, occasional small deviations between the calculated and measured forces are observed, arising from pinning events that cannot be easily captured experimentally, as pinned points may lie outside of view.

The bottom inferred force $F_b^{inf}$ and calculated force $F_b^{calc}$ still agree within error (Figs. 3m–o). However, it is noticeable that agreement is poorer for the bottom than the top of the bridge. This is to some extent expected since the comparison is less direct than at the top surface. On Glass and DMS, $F_b^{inf}$ and $F_b^{calc}$ show good agreement, with an overall behavior of simply shifted version from the top surface data. On LIS, however, a noticeable difference is observed between the inferred and calculated force (see dashed square in Fig. 3o), and several factors contribute to this complex comparison. First, the exerted forces on LIS are considerably smaller than those on Glass or DMS, making the relative errors inevitably larger (Figs. 3m–o). Second, there are uncertainties in the surface tension used to calculate $F_b^{calc}$ (Eq. 3), as the capillary bridge is likely to be cloaked by the LIS lubricant. Cloaking is a well-known phenomenon in LIS.[46–48] Here, we estimate the spreading coefficient of the lubricant over the capillary bridge to be $S \sim 20$ mN/m, suggesting full cloaking. Determining an effective surface tension for the cloaked capillary bridge is not straightforward as the surface tension of a thin film is known to vary with its thickness.[49] In this setup, we are unable to measure the thickness of the cloaking film which may not be uniform and may evolve over the course of an experiment, and lubricant transport between the two surfaces through the capillary bridge is possible. To reflect this uncertainty, we adopt an effective droplet–gas surface tension of $\gamma_{dg} = 50 \pm 2 \; mN \, m^{-1}$, obtained by averaging our pendant drop measurement (analyzed by the Opendrop software[50,51]) with a value inferred from literature data for a similar system[52] (see Supporting Information SI 10 for details). While this reduction from the uncloaked droplet's surface tension ($\sim 67 \; mN \, m^{-1}$) to the effective cloaked value ($\sim 50 \; mN \, m^{-1}$) is relatively small, it is sufficient to significantly affect the data considering the small forces at play ($\sim 20$–25% of the force value). Finally, the presence of a lubricant ridge around the capillary bridge on LIS further complicates the measurement of the meridional radius of curvature $R_2$, which is required for calculating $F_b^{calc}$ (see Supporting Information SI 11 for more details).

The above observations and the analysis of the bridges' geometrical features and exerted forces raise three immediate questions: i) For solid surfaces with roughness or chemical heterogeneities, how can the stick–slip motion of the contact line be described, and what is its impact on geometry and force? ii) Why does LIS show poorer bottom force comparison between inferred and calculated force compared to Glass and DMS surfaces? iii) Both DMS and LIS are hydrophobic surfaces but have distinct capillary behaviors. If combined in a single capillary bridge, which behavior would dominate?

**Simulation of stick–slip motion on heterogenous surfaces**

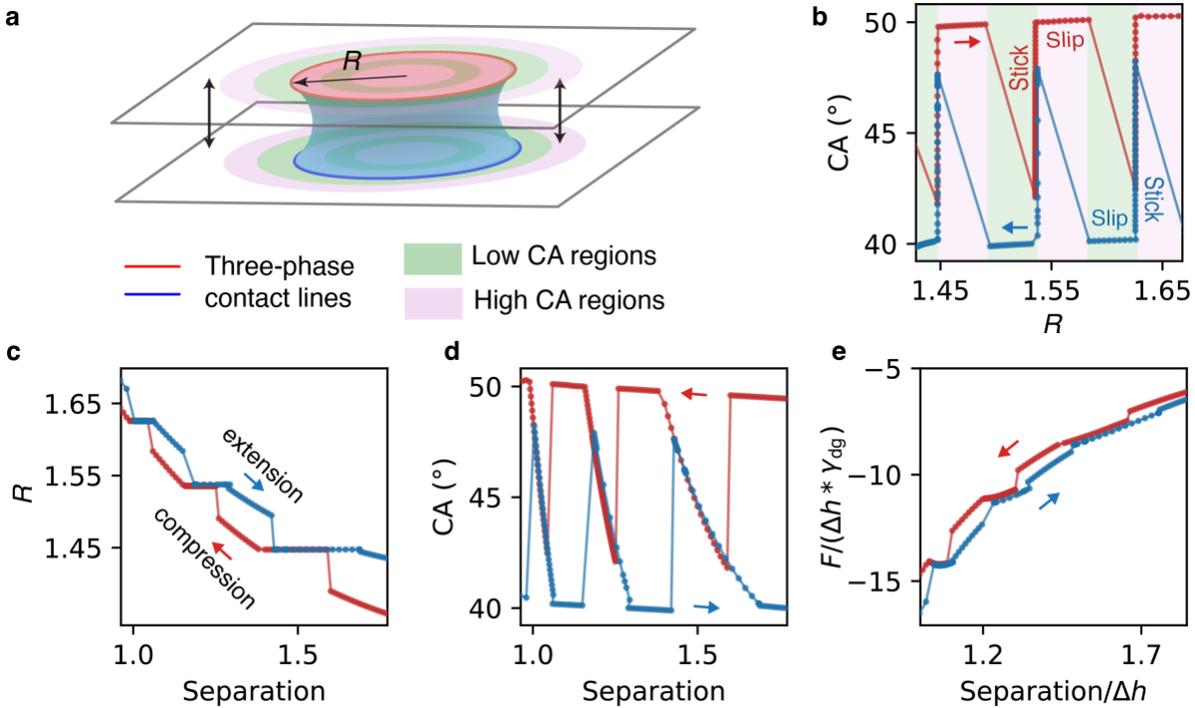

**Figure 4**. Simulated capillary bridge behavior on a binary-patterned surface featuring alternating high-CA (purple, 50°) and low-CA (green, 40°) rings. The three-phase contact line between the bridge, air, and solid are highlighted in red and blue in (a). The motion of the contact line across the heterogeneous surface exhibits stick–slip behavior, with the corresponding geometric parameters (contact radius $R$, and contact angle CA measured near the contact line) and capillary forces shown in (b–e). Gravitational effects are neglected to ensure symmetric contact with both patterned plates, resulting in equal top and bottom contact radii. $\Delta h$ denotes the maximum plate separation during the compression–extension cycle, and $\gamma_{dg}$ is the droplet–gas interfacial tension. Because the correlation between bridge separation and base radius is nonlinear, (b) additionally shows the CA plotted against $R$ to directly reflect the imposed binary pattern, whereas (c–e) present all results as a function of separation for consistent comparison with Fig. 3.

Experimentally, investigating the stick–slip motion in capillary bridges is challenging due to the asymmetric and highly localized nature of contact line pinning. To gain better insights into the underlying mechanisms, we performed numerical simulations examining the bridge geometry evolution and capillary force dynamics during compression and extension cycles over chemically heterogeneous surfaces. Practically, we use a binary-patterned substrate featuring alternating regions of high and low contact angles (50° and 40°, respectively) to systematically model surface chemical heterogeneity (Fig. 4a). For simplicity, we ignore gravity in the simulations since it is not critical for the contact line pinning-depinning behavior. During compression (Fig. 4b), the three-phase contact line advances until it encounters a boundary transitioning from low-CA to high-CA regions, where strong pinning occurs. At this stage, the contact radius $R$ remains fixed while the bridge height continues to decrease under the applied compression, resulting in an increase in the measured CA. Once the local contact angle reaches the prescribed high-CA value, the pinning constraint is released, allowing the contact line to advance across the high-CA region. When reaching the subsequent low-CA region, the contact line exhibits rapid forward motion due to the energetically favorable wetting conditions. This alternating sequence of pinning and release events repeats throughout the compression process. During stretching, the process is reversed. the receding contact line becomes preferentially pinned at boundaries transitioning toward low-CA regions, where higher wettability disadvantages receding. This pinning behavior during both compression and extension cycles generates distinctive stepwise variations and hysteresis both in the $R$ and CA evolutions (Figs. 4c–d), providing clear experimental signatures of the stick–slip phenomenon.

The capillary force in the above simulation was calculated by Eq. 1 with the pressure obtained from the simulation model and normalized by the product of the plate separation change $\Delta h$ and the droplet–gas interfacial tension $\gamma_{dg}$. The resulting force (Fig. 4e) exhibits a stepwise behavior similar to that visible for Glass in Fig. 3j. Similarly, it is possible to simulate roughness-induced contact line pinning and depinning, also inducing stepwise features in force and geometry measurements (see details in Supporting Information SI 12). These results confirm that ability of simulations to capture the fundamental aspects of the stick–slip hysteresis on binary-patterned and rough surfaces, and offer a basis for studying more complex substrate designs and interfacial interactions.

**Top and bottom symmetry of the capillary bridges**

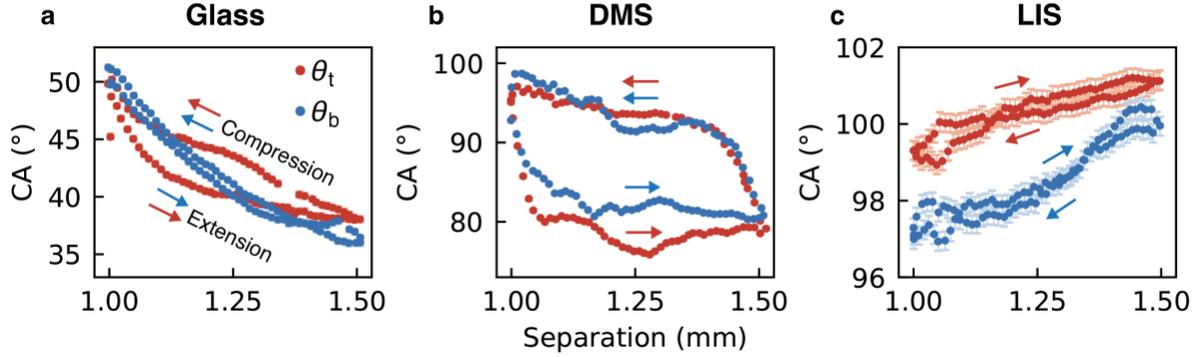

**Figure 5.** Variations in the capillary bridges' top contact angle $\theta_t$ (red) and bottom contact angle $\theta_b$ (blue) at their contact with Glass (a), DMS (b), and LIS (c). Error bars represent two standard errors and may not be visible on Glass or DMS. The data shown comes from a different set than that presented in Fig. 3 but is obtained following an identical protocol.

As discussed in section 3.1, the discrepancy between the bottom inferred force $F_b^{\text{inf}}$ and calculated force $F_b^{\text{calc}}$ of LIS can be attributed to uncertainty in the parameters used to obtain $F_b^{\text{calc}}$ (Eq. 3). For capillary bridge on LIS, the capillary shape can be distorted due to differences in the top and bottom lubricant menisci that can modify local interfacial stresses. Also, the bridge–air interface may be non-ideal due to the presence of a cloaking lubricant film. Consequently, $F_b^{\text{calc}}$ is a local approximation that may not capture the true force in asymmetric or complex interfaces.

To examine such symmetry effects, we analyze the evolution of the CAs at the top and bottom surfaces for Glass, DMS, and LIS systems (Fig. 5). On Glass, the top and bottom CA show similar monotonic decrease upon extension, consistent with contact line pinning and stick–slip motion (Fig. 5a). On DMS, both surfaces exhibit the characteristic hysteresis loop discussed in section 3.1 (Fig. 5b). For LIS, the CAs on both surfaces increase by $2 - 3°$ during extension with angles overlapping during outward and return motions, indicating the absence of hysteresis. As the capillary bridge is stretched on LIS, the droplet Laplace pressure decreases, explaining the increase in the CA.

Notably, a small but consistent offset ($\sim 2°$) exists between the top and bottom CAs, $\theta_t$ and $\theta_b$ (Fig. 5c). Unlike pinning-induced asymmetry, this offset originates from the pressure ratio between the bridge and the lubricant.[53,54] For the typical capillary bridge considered in this work, the hydrostatic pressure difference between the top and bottom of the bridge is sufficient to account for the observed CA asymmetry. The maximum capillary force on LIS is around 5 to 10 times lower than for DMS and Glass (Figs. 3j–o), making the geometry more sensitive to gravitational effects. The maximum bottom force for LIS is ∼ 0.15 mN (Fig. 3o), while the gravity term $\rho g V$ (Eq. 4) for such capillary bridge is ∼ 0.12 mN, indicating gravity and surface tension effects become comparable. In contrast, capillary bridges on Glass and DMS formed by the same droplet reach maximum forces of $0.7 - 2.2$ mN (Figs. 3m, 3n), where surface tension remains dominant throughout most of the extension-compression cycle.

**Asymmetric hydrophobic capillary bridge with DMS and LIS by design**

So far, our results have focused on capillary bridges formed between identical top and bottom surfaces. Moving from Glass to DMS to LIS, the surfaces become progressively more hydrophobic, with increasing CAs. Glass and DMS represent widely used bare or functionalized solid surfaces, where the capillary bridge behavior is dominated by contact line pinning. LIS, in contrast, exhibits distinct behavior arising from the lubricant's fluid nature, characterized by low friction, dynamic menisci, and low exerted forces. The capillary bridge on LIS is not perfectly symmetric at the top and bottom, although the effects of asymmetry are generally subtle. To explore this further, we design experiments with deliberately asymmetric systems, using different surfaces at the top and bottom. To prevent one surface from dominating, it is helpful to retain some similarity by selecting surfaces of comparable hydrophobicity. Here, we do this by using DMS and LIS surfaces, with each surface alternately positioned on the top or bottom. This asymmetric system is interesting because it represents two hydrophobic surfaces: one a solid surface that exhibits typical hysteresis and pinning, and the other a liquid-infused surface which is smooth, frictionless, and dynamically adaptive.

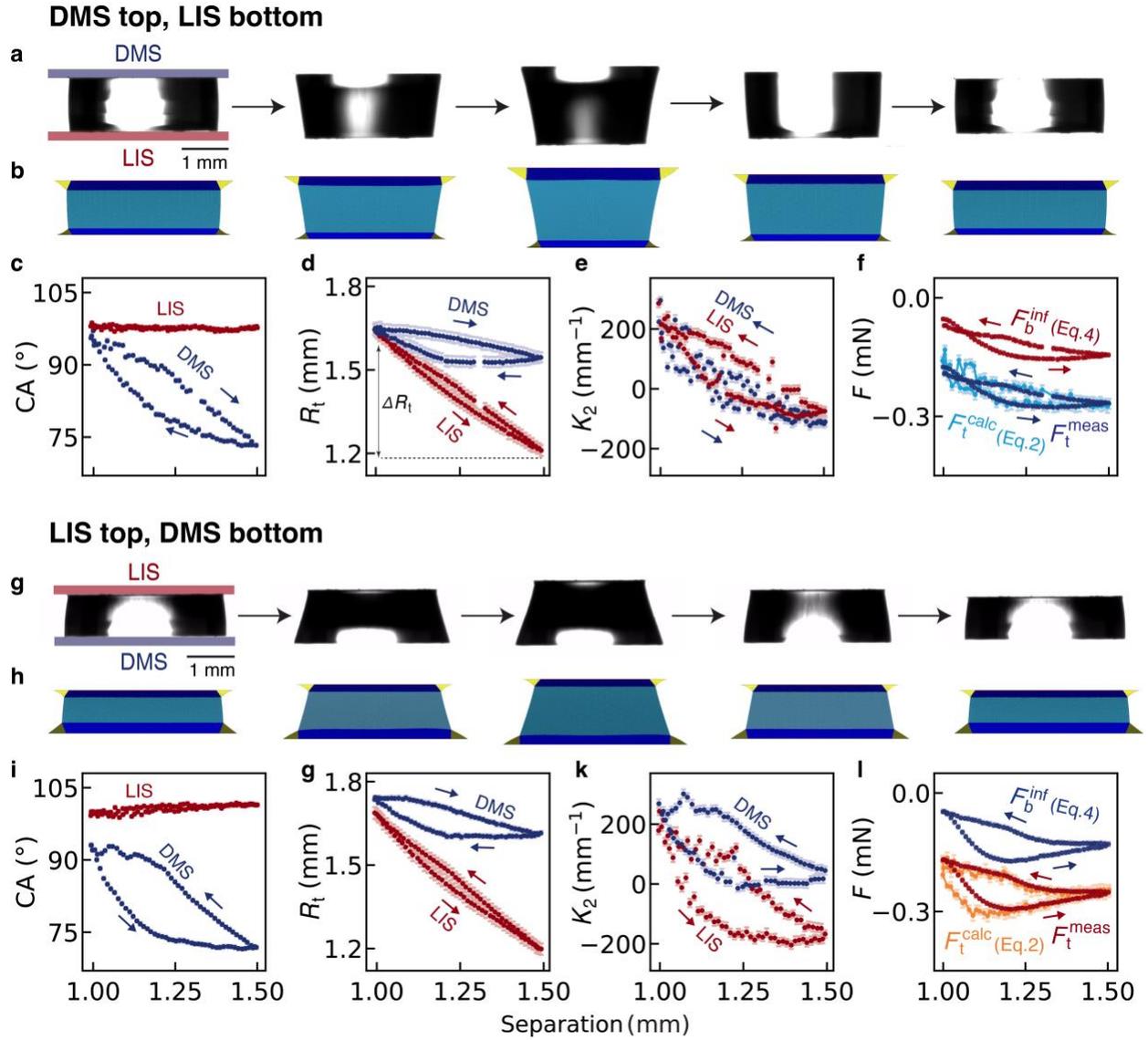

**Figure 6.** Comparative behavior of asymmetric capillary bridge between LIS (red traces) and DMS (blue traces). (a, g) Experimental images of bridge configurations during extension-compression. (b, h) Computational simulations incorporating the lubricant ridge (see the Methods section) are performed for both LIS and DMS, accounting for lubricant transport. The contact angles (c, i) remain almost constant on LIS, whereas DMS exhibits a pronounced hysteretic behavior, exploring a larger range of angles than that observed in the symmetric DMS system in Fig. 3e. The evolution of capillary bridge–surface top contact radius $R_t$ (d, g), meridional curvature $K_2$ (e, k), and the exerted forces (f, l) are shown. Change in contact radius, denoted by $\Delta R_t$, is marked as an example in (d). Error bars represent standard errors.

Fig. 6 shows the results of designed asymmetric capillary bridge. The CA measured on the LIS remains almost constant at $98 - 100°$, regardless of whether LIS is placed at the top or bottom (Figs. 6c, 6i, red traces). The small difference in CA between the two configurations arises from the gravitational deformation of the bridge, consistent with the behavior observed in the symmetric system (Fig. 5c). However, the evolution of the CA on the DMS surfaces is markedly different from that observed in the symmetric DMS system. In the symmetric DMS case, pinning induces a CA hysteresis loop with two plateaus, an advancing angle of ~98° and a receding angle of ~84° (Figs. 3e, 5b). In the asymmetric case here, this hysteresis loop becomes elongated. The CA changes monotonically with larger absolute variations (Figs. 6c, 6i, blue traces), and without stabilizing at the typical advancing or receding plateaus. This behavior originates from two concurrent effects. First, the surface tension of the capillary bridge is altered because the bridge is cloaked by the lubricant. Second, lubricant is transferred from the LIS to the DMS surface. This lubricant accumulation forms a ridge at the bridge–DMS contact line, which broadens the accessible range of CAs. To probe this effect independently, we perform droplet experiments on DMS under cycles of volume variation as lubricant diffuses slowly towards it (see SI 8 and Video S1). Initially, we can identify advancing and receding angles in agreement with Figs. 3e and 5b for the capillary bridge setup, as the lubricant has no or very limited contact with the droplet. As the lubricant progressively wets the droplets, a ridge forms at the contact line on DMS. This ridge pins the droplet and produces larger droplet CA variation than observed before lubricant contact. To further verify that lubricant can migrate from the LIS to the DMS during capillary bridge deformation, we perform an experiment using a dyed silicone-oil-infused LIS (bottom) and a DMS surface (top). During bridge compression, we can observe the dyed lubricant migrate along the bridge to the DMS surface (see details in Supporting Information SI 13), providing direct qualitative evidence of lubricant transfer.

The lubricant ridge is too small for its details to be directly resolved with our experimental setup, but its role can be assessed through computational simulations. This is achieved by introducing an oil meniscus around the three-phase contact region at both top and bottom surfaces. For the lubricant ridge on DMS, the oil–gas contact angle $\theta_{og}$ is set at 35° (inferred from Fig. S5), while for LIS it is set to 15° (representing high spreading; the resulting behavior is very similar when lower contact angle is employed). Other relevant interfacial tensions are taken from measurements and derivation (see Methods section 2.4 and Eq. 5). To maintain symmetry and to balance computational costs and accuracy, the ridge volumes used in the simulations are larger than those in the experiments. This approximation is valid as long as the ridge remains much smaller than the capillary bridge itself, since the local Neumann balance at the three-phase contact line is preserved. As shown in Figs. 6a–b and 6g–h, the simulated bridge geometries closely match the experimental images, confirming the importance of including lubricant ridges at both surfaces. Furthermore, the force calculated from the simulations using Eq. 6 agrees well with experimental measurements (Supporting Information SI 14). Overall, the consistency in both geometry and force demonstrates that the model in Fig. 2 reliably captures the physics of capillary bridges involving lubricant ridges. Beyond the present application, this framework can be extended to describe liquid–liquid, liquid–solid, and three-phase interactions on functional surfaces.

We now further examine the geometrical and force responses in these two asymmetric systems. The radius of the contact area changes by $\Delta R_t = 0.5$ mm at the LIS interface with no hysteresis, and by $\Delta R_t < 0.1$ mm at the DMS interface with some hysteresis (Figs. 6d, 6g). As expected, the bridge preferentially slides across LIS compared to DMS. When LIS is the bottom surface (with DMS on top), the meridional curvature $K_2$ remains comparable near the top and bottom of the bridge (Fig. 6e). This is expected, since the CAs are similar at the onset of the extension (~95° for DMS and ~98° for LIS). However, when LIS is positioned at the top and DMS at the bottom, the $K_2$ at the two surfaces no longer match. An offset and distinct ranges of curvature values are observed near the top and bottom surfaces (Fig. 6k). This effect can be explained by the influence of gravity on the capillary bridge geometry (CA and $R_t$), amplified by enhanced lubricant transfer from LIS when it is on top.

Comparison of the measured, calculated and inferred capillary force shows consistently good agreement (Figs. 6f, 6l), with larger hysteresis observed for the LIS-top-DMS-bottom system. In such system, gravity promotes oil transfer from the LIS top to the DMS bottom, leading to the formation of a large oil ridge that enhances pinning or friction during extension or compression of the bridge. Overall, the variations in force are comparable to those in the symmetric LIS systems (Figs. 3l, 3o) regardless of the configuration, which can be explained by a combination of two factors. First, when LIS is present in the system, the surface tension of the bridge drops from $\sim 67\ mN/m^{55}$ to $\sim 50\ mN/m$ due to cloaking. Second, the LIS offers a non-pinning surface, allowing contact line to move preferentially and minimize the energy required to extend the bridge.

Consequently, contact angles and contact radii follow similar trends on each surface in symmetric systems, whereas the radii of curvature and exerted force reveal the effects of oil transfer and friction characteristic of asymmetric bridges.

## Conclusion

In this study, we systematically investigate capillary bridges on LIS and compare the observed behavior with two 'standard' non-infused solid surfaces: hydrophilic Glass and hydrophobic DMS. The good agreement between experiments, modelling, and theory demonstrates that our model accurately captures the behavior of capillary bridges in the quasistatic limit, including for LIS through the use of an apparent contact angle. In agreement with previous studies, contact line pinning is prevalent on Glass, giving rise to a complicated stick–slip motion; DMS exhibits typical contact angle hysteresis on hydrophobic surfaces, with well-defined geometrical features during extension-compression.[23,24,27,56] In contrast, LIS [33,48,57,58] exhibits markedly distinct trends due to the absence of pinning. First, no hysteresis is observed during bridge extension or compression. Second, the variation in capillary force is substantially reduced thanks to the frictionless nature of the lubricant. Third, the small forces and absence of pinning allow gravity to break the bridge symmetry, an effect that is often masked by pinning on solid surfaces. On LIS, gravity can affect the apparent CA by altering the pressure balance within the capillary, in agreement with theoretical predictions.[35,53] Finally, lubricant cloaking on LIS[46–48] reduces the effective surface tension and allows lubricant transport between the surfaces. The effect is most pronounced in asymmetric capillary bridges formed by a LIS and DMS, where lubricant transfer produces a ridge on the DMS surface, modifying capillary interactions and introducing localized pinning.

Further work will focus on dynamic interactions between the lubricant and capillary bridges, particularly the evolution of oil ridges over time under mechanical deformation and varying pressures. Incorporating dynamic effects, such as lubricant viscosity and the velocity of capillary bridge extension-compression, would extend the predictive capability of the model to practical applications, including printing, coatings, cell culture, and microfluidics. Overall, this study establishes a fundamental framework that can help design functional liquid-like surfaces with tunable and controllable capillary interactions.

## Data Availability

All the data and videos used to create the figures presented in this paper as well as example simulation scripts are freely available at the Durham Research Online Repository: https://collections.durham.ac.uk/files/r26m311p40q ( doi:10.15128/r26m311p40q )

## Supporting Information

Details on surface characterization, experimental setup and imaging procedures, droplet placement, capillary bridge focusing and geometry analysis, reproducibility of measurements, gravitational effects, oil ridge formation and induced pinning on DMS, contact line pinning, effective surface tension for cloaked systems, fitting of bridges with oil ridges, lubricant transfer, and comparisons between experiments and simulations (PDF).

Oil ridge formation and induced pinning on DMS (AVI).

## Corresponding Author

kislon.voitchovsky@durham.ac.uk

halim.kusumaatmaja@ed.ac.uk

## Author Contributions

SJG and KS contributed equally. SJG performed the experiments with inputs from KS. KS performed the simulations. SJG and KS analyzed the results with help from KV and HK. The manuscript was written through contributions of all authors.

## Funding Sources

This research is supported by EPSRC Soft Matter and Functional Interfaces (SOFI, Grant EP/L015536/1), EPSRC Soft Matter for Formulation and Industrial Innovation (SOFI2, Grant EP/S023631/1), EPSRC Fellowships (Grant EP/V034154/2, Grant EP/S028234/1), and Leverhulme Trust (Grant RPG-2022-140).

## Acknowledgment

SJG thanks help from Lisong Yang and Colin Bain for DMS surface preparation (including the silanization protocol), Gary Wells and Glen McHale for LIS surface preparation. KS thanks Alex Brown and Alvin C. M. Shek for discussion on simulation. We thank the electronic and mechanical

**References**

(1) Li, M.; Shi, L.; Wang, X. Physical Mechanisms behind the Wet Adhesion: From Amphibian Toe-Pad to Biomimetics. *Colloids Surf., B* **2021**, *199*, 111531.
(2) Tan, D.; Zhu, B.; Xiao, K.; Li, L.; Shi, Z.; Liu, Q.; Gorb, S.; Gao, H.; Pham, J. T.; Liu, Z.; Xue, L. Nanosized Contact Enables Faster, Stronger, and Liquid-Saving Capillary Adhesion. *ACS Nano* **2025**, *19* (9), 8571–8578.
(3) Hornbaker, D. J.; Albert, R.; Albert, I.; Barabási, A.-L.; Schiffer, P. What Keeps Sandcastles Standing? *Nature* **1997**, *387* (6635), 765–765.
(4) Miot, M.; Veylon, G.; Wautier, A.; Philippe, P.; Nicot, F.; Jamin, F. Numerical Analysis of Capillary Bridges and Coalescence in a Triplet of Spheres. *Granular Matter* **2021**, *23* (3), 1–18.
(5) Liu, Y.; Che, P.; Zhang, B.; Yang, J.; Gao, H.; Feng, J.; Wu, Y.; Jiang, L. One-Step Patterning of Organic Semiconductors on Gold Electrodes via Capillary-Bridge Manipulation. *ACS Appl. Mater. Interfaces* **2022**, *14* (28), 32761–32770.
(6) Saadat, M.; Yang, J.; Dudek, M.; Øye, G.; Tsai, P. A. Microfluidic Investigation of Enhanced Oil Recovery: The Effect of Aqueous Floods and Network Wettability. *J. Pet. Sci. Eng.* **2021**, *203*, 108647.
(7) Zhang, P.; Chen, Z.; Brown, K. G.; Meeussen, J. C. L.; Gruber, C.; Garrabrants, A. C.; Kosson, D. S. Drying Model of a High Salt Content Cementitious Waste Form: Effect of Capillary Forces and Salt Solution. *Cem. Concr. Res.* **2021**, *146*, 106459.
(8) Bian, S.; Tai, C.-F.; Halpern, D.; Zheng, Y.; Grotberg, J. B. Experimental Study of Flow Fields in an Airway Closure Model. *J. Fluid Mech.* **2010**, *647*, 391–402.
(9) Gabrielsson, E. O.; Jung, Y. H.; Han, J. H.; Joe, D. J.; Simon, D. T.; Lee, K. J.; Berggren, M. Autonomous Microcapillary Drug Delivery System Self-Powered by a Flexible Energy Harvester. *Adv. Mater. Technol.* **2021**, *6* (11), 2100526.
(10) Pakarinen, O. H.; Foster, A. S.; Paajanen, M.; Kalinainen, T.; Katainen, J.; Makkonen, I.; Lahtinen, J.; Nieminen, R. M. Towards an Accurate Description of the Capillary Force in Nanoparticle-Surface Interactions. *Modelling Simul. Mater. Sci. Eng.* **2005**, *13* (7), 1175–1186.
(11) Butt, H.-J.; Kappl, M. Normal Capillary Forces. *Adv. Colloid Interface Sci.* **2009**, *146* (1), 48–60.
(12) Berthier, J. Theory of Wetting. In *Micro-Drops and Digital Microfluidics (Second Edition)*; Berthier, J., Ed.; William Andrew Publishing, 2013; pp 7–73.
(13) Fan, M.; Fan, Z.; Xu, Z.; Li, J.; Li, C.; Yang, Z. Investigation of Capillary Forces and Capillary Bridges between an End-Adjusted Three-Finger Microgripper with Hydrophobic Side Surface and a Plate. *J. Adhes. Sci. Technol.* **2024**, *38* (10), 1702–1717.
(14) Rodrigues, M. S.; Coelho, R. C. V.; Teixeira, P. I. C. Dynamics of Liquid Bridges between Patterned Surfaces. *Phys. D* **2024**, *469*, 134322.
(15) Lee, E.; Müller-Plathe, F. Contact Line Friction and Dynamic Contact Angles of a Capillary Bridge between Superhydrophobic Nanostructured Surfaces. *J. Chem. Phys.* **2022**, *157* (2), 024701.

(16) Daniel, D.; Lay, C. L.; Sng, A.; Jun Lee, C. J.; Jin Neo, D. C.; Ling, X. Y.; Tomczak, N. Mapping Micrometer-Scale Wetting Properties of Superhydrophobic Surfaces. *Proc. Natl. Acad. Sci. U.S.A.* **2019**, *116* (50), 25008–25012.

(17) Liimatainen, V.; Vuckovac, M.; Jokinen, V.; Sariola, V.; Hokkanen, M. J.; Zhou, Q.; Ras, R. H. A. Mapping Microscale Wetting Variations on Biological and Synthetic Water-Repellent Surfaces. *Nat. Commun.* **2017**, *8* (1), 1798.

(18) Cassin, F.; Hahury, R.; Lançon, T.; Franklin, S.; Weber, B. The Nucleation, Growth, and Adhesion of Water Bridges in Sliding Nano-Contacts. *J. Chem. Phys.* **2023**, *158* (22), 224703.

(19) Cheng, S.; Robbins, M. O. Capillary Adhesion at the Nanometer Scale. *Phys. Rev. E* **2014**, *89* (6), 062402.

(20) Nguyen, N. N.; Davani, S.; Asmatulu, R.; Kappl, M.; Berger, R.; Butt, H.-J. Nano-Capillary Bridges Control the Adhesion of Ice: Implications for Anti-Icing via Superhydrophobic Coatings. *ACS Appl. Nano Mater.* **2022**, *5* (12), 19017–19024.

(21) Tanaka, K.; Ito, T.; Nishiyama, Y.; Fukuchi, E.; Fuchiwaki, O. Double-Nozzle Capillary Force Gripper for Cubical, Triangular Prismatic, and Helical 1-Mm-Sized-Objects. *IEEE Robot. Autom. Lett.* **2022**, *7* (2), 1324–1331.

(22) Portuguez, E.; Alzina, A.; Michaud, P.; Hourlier, D.; Smith, A. Study of the Contact and the Evaporation Kinetics of a Thin Water Liquid Bridge between Two Hydrophobic Plates. *Adv. Mater. Phys. Chem.* **2017**, *7* (4), 99–112.

(23) Shi, Z.; Zhang, Y.; Liu, M.; Hanaor, D. A. H.; Gan, Y. Dynamic Contact Angle Hysteresis in Liquid Bridges. *Colloids Surf., A* **2018**, *555*, 365–371.

(24) De Souza, E. J.; Gao, L.; McCarthy, T. J.; Arzt, E.; Crosby, A. J. Effect of Contact Angle Hysteresis on the Measurement of Capillary Forces. *Langmuir* **2008**, *24* (4), 1391–1396.

(25) Yang, L.; Sega, M.; Harting, J. Capillary-Bridge Forces between Solid Particles: Insights from Lattice Boltzmann Simulations. *AIChE J.* **2021**, *67* (9), e17350.

(26) Odunsi, M. S.; Morris, J. F.; Shattuck, M. D. Hysteretic Behavior of Capillary Bridges between Flat Plates. *Langmuir* **2023**, *39* (37), 13149–13157.

(27) Chen, H.; Amirfazli, A.; Tang, T. Modeling Liquid Bridge between Surfaces with Contact Angle Hysteresis. *Langmuir* **2013**, *29* (10), 3310–3319.

(28) Li, J.; Ueda, E.; Paulssen, D.; Levkin, P. A. Slippery Lubricant-Infused Surfaces: Properties and Emerging Applications. *Adv. Funct. Mater.* **2019**, *29* (4), 1802317.

(29) Villegas, M.; Zhang, Y.; Abu Jarad, N.; Soleymani, L.; Didar, T. F. Liquid-Infused Surfaces: A Review of Theory, Design, and Applications. *ACS Nano* **2019**, *13* (8), 8517–8536.

(30) Long, Y.; Yin, X.; Mu, P.; Wang, Q.; Hu, J.; Li, J. Slippery Liquid-Infused Porous Surface (SLIPS) with Superior Liquid Repellency, Anti-Corrosion, Anti-Icing and Intensified Durability for Protecting Substrates. *Chem. Eng. J.* **2020**, *401*, 126137.

(31) Dawson, J.; Coaster, S.; Han, R.; Gausden, J.; Liu, H.; McHale, G.; Chen, J. Dynamics of Droplets Impacting on Aerogel, Liquid Infused, and Liquid-Like Solid Surfaces. *ACS Appl. Mater. Interfaces* **2023**, *15* (1), 2301–2312.

(32) Zhu, Y.; McHale, G.; Dawson, J.; Armstrong, S.; Wells, G.; Han, R.; Liu, H.; Vollmer, W.; Stoodley, P.; Jakubovics, N.; Chen, J. Slippery Liquid-Like Solid Surfaces with Promising Antibiofilm Performance under Both Static and Flow Conditions. *ACS Appl. Mater. Interfaces* **2022**, *14* (5), 6307–6319.

(33) Wong, T. S.; Kang, S. H.; Tang, S. K. Y.; Smythe, E. J.; Hatton, B. D.; Grinthal, A.; Aizenberg, J. Bioinspired Self-Repairing Slippery Surfaces with Pressure-Stable Omniphobicity. *Nature* **2011**, *477* (7365), 443–447.

(34) Hardt, S.; McHale, G. Flow and Drop Transport Along Liquid-Infused Surfaces. *Annu. Rev. Fluid Mech.* **2022**, *54* (1), 83–104.

(35) Shek, A. C. M.; Semprebon, C.; Panter, J. R.; Kusumaatmaja, H. Capillary Bridges on Liquid-Infused Surfaces. *Langmuir* **2021**, *37* (2), 908–917.

(36) Deng, R.; Yang, L.; Bain, C. D. Combining Inkjet Printing with Emulsion Solvent Evaporation to Pattern Polymeric Particles. *ACS Appl. Mater. Interfaces* **2018**, *10* (15), 12317–12322.

(37) Orme, B. V.; McHale, G.; Ledesma-Aguilar, R.; Wells, G. G. Droplet Retention and Shedding on Slippery Substrates. *Langmuir* **2019**, *35* (28), 9146–9151.

(38) Goodband, S. J.; Armstrong, S.; Kusumaatmaja, H.; Voïtchovsky, K. Effect of Ageing on the Structure and Properties of Model Liquid-Infused Surfaces. *Langmuir* **2020**, *36* (13), 3461–3470.

(39) Goodband, S.; Kusumaatmaja, H.; Voïtchovsky, K. Development of a Setup to Characterize Capillary Liquid Bridges between Liquid Infused Surfaces. *AIP Adv.* **2022**, *12*, 015120.

(40) Moreno-Labella, J.; Munoz-Martin, D.; Márquez, A.; Morales, M.; Molpeceres, C. Numerical Study of Water-Glycerol BA-LIFT: Analysis and Simulation of Secondary Effects. *Opt. Laser Technol.* **2021**, *135*, 106695.

(41) Brakke, K. A. The Surface Evolver. *Exp. Math.* **1992**, *1* (2), 141–165.

(42) Bowen, J.; Cheneler, D. Closed-Form Expressions for Contact Angle Hysteresis: Capillary Bridges between Parallel Platens. *Colloids Interfaces* **2020**, *4* (1), 13.

(43) Kusumaatmaja, H.; Yeomans, J. M. Modeling Contact Angle Hysteresis on Chemically Patterned and Superhydrophobic Surfaces. *Langmuir* **2007**, *23* (11), 6019–6032.

(44) Abbas, A.; Wells, G. G.; McHale, G.; Sefiane, K.; Orejon, D. Silicone Oil-Grafted Low-Hysteresis Water-Repellent Surfaces. *ACS Appl. Mater. Interfaces* **2023**, *15* (8), 11281–11295.

(45) Cao, Y.; Jana, S.; Tan, X.; Bowen, L.; Zhu, Y.; Dawson, J.; Han, R.; Exton, J.; Liu, H.; McHale, G.; Jakubovics, N. S.; Chen, J. Antiwetting and Antifouling Performances of Different Lubricant-Infused Slippery Surfaces. *Langmuir* **2020**, *36* (45), 13396–13407.

(46) Günay, A. A.; Sett, S.; Ge, Q.; Zhang, T.; Miljkovic, N. Cloaking Dynamics on Lubricant-Infused Surfaces. *Adv. Mater. Interfaces* **2020**, *7* (19), 2000983.

(47) Schellenberger, F.; Xie, J.; Encinas, N.; Hardy, A.; Klapper, M.; Papadopoulos, P.; Butt, H.-J.; Vollmer, D. Direct Observation of Drops on Slippery Lubricant-Infused Surfaces. *Soft Matter* **2015**, *11* (38), 7617–7626.

(48) Smith, J. D.; Dhiman, R.; Anand, S.; Reza-Garduno, E.; Cohen, R. E.; McKinley, G. H.; Varanasi, K. K. Droplet Mobility on Lubricant-Impregnated Surfaces. *Soft Matter* **2013**, *9* (6), 1772–1780.

(49) Pepper, K. G.; Bahrim, C.; Tadmor, R. Interfacial Tension and Spreading Coefficient of Thin Films: Review and Future Directions. *J. Adhes. Sci. Technol.* **2011**, *25* (12), 1379–1391.

(50) Berry, J. D.; Neeson, M. J.; Dagastine, R. R.; Chan, D. Y. C.; Tabor, R. F. Measurement of Surface and Interfacial Tension Using Pendant Drop Tensiometry. *J. Colloid Interface Sci.* **2015**, *454*, 226–237.

(51) Daerr, A.; Mogne, A. Pendent_Drop: An ImageJ Plugin to Measure the Surface Tension from an Image of a Pendent Drop. *J. Open Res. Softw.* **2016**, *4* (1), e3.

(52) Gunjan, M. R.; Kumar, A.; Raj, R. Cloaked Droplets on Lubricant-Infused Surfaces: Union of Constant Mean Curvature Interfaces Dictated by Thin-Film Tension. *Langmuir* **2021**, *37* (22), 6601–6612.
(53) Semprebon, C.; McHale, G.; Kusumaatmaja, H. Apparent Contact Angle and Contact Angle Hysteresis on Liquid Infused Surfaces. *Soft Matter* **2016**, *13* (1), 101–110.
(54) Semprebon, C.; Sadullah, M. S.; McHale, G.; Kusumaatmaja, H. Apparent Contact Angle of Drops on Liquid Infused Surfaces: Geometric Interpretation. *Soft Matter* **2021**, *17* (42), 9553–9559.
(55) Takamura, K.; Fischer, H.; Morrow, N. R. Physical Properties of Aqueous Glycerol Solutions. *J. Pet. Sci. Eng.* **2012**, *98–99*, 50–60.
(56) De Souza, E. J.; Brinkmann, M.; Mohrdieck, C.; Crosby, A.; Arzt, E. Capillary Forces between Chemically Different Substrates. *Langmuir* **2008**, *24* (18), 10161–10168.
(57) Sadullah, M. S.; Panter, J. R.; Kusumaatmaja, H. Factors Controlling the Pinning Force of Liquid Droplets on Liquid Infused Surfaces. *Soft Matter* **2020**, *16* (35), 8114–8121.
(58) Lafuma, A.; Quéré, D. Slippery Pre-Suffused Surfaces. *Europhys. Lett.* **2011**, *96* (5), 56001.

Supporting Information

# Stretching and Compressing Capillary Bridges on Hydrophilic, Hydrophobic, and Liquid-infused Surfaces

Sarah Jane Goodband[1,⊥], Ke Sun[1,⊥], Kislon Voïtchovsky[1*], Halim Kusumaatmaja[2*]

[1]Department of Physics, Durham University, Durham DH1 3LE, UK

[2]Institute for Multiscale Thermofluids, School of Engineering, The University of Edinburgh, Edinburgh, EH9 3FB, UK

[⊥] These authors contribute equally

Correspondence:

* Email: kislon.voitchovsky@durham.ac.uk

* Email: halim.kusumaatmaja@ed.ac.uk

**Table of contents**

# 1. Surface topography characterization using atomic force microscopy (AFM)

**Table S1**. AFM topographic characterization of glass and DMS surfaces. Imaging is performed in amplitude modulation using a Cypher ES AFM (Oxford Instruments, USA) at 25 °C in pure water. The AFM tip/cantilever (SNL-10 tip A, Bruker AFM Probes, USA, nominal spring constant of 0.43 N/m) was calibrated using its thermal spectrum after the measurements. Representative AFM images and the measured values (average ± standard deviation) are shown, obtained from three repeated measurements across randomly selected areas of each sample. The roughness statistics are processed using Gwyddion[1].

| Material | AFM Topographic image | Image size | Mean roughness ($S_a$) in nm | RMS roughness ($S_q$) in nm | Surface area /projected area |
|---|---|---|---|---|---|
| Glass | 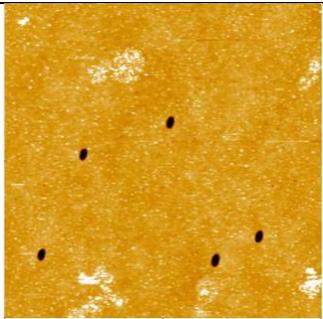 | XY: 10 μm<br><br>Z: 10 nm (colour scale range) | 0.449 ± 0.027 | 1.031 ± 0.207 | 1.00039 ± 0.00008 |
| DMS | 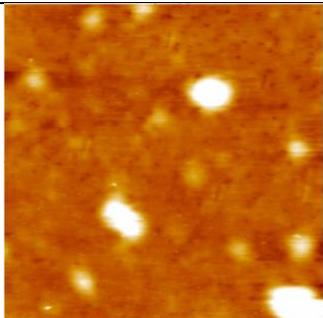 | XY: 10 μm<br><br>Z: 30 nm (colour scale range) | 2.297 ± 0.221 | 3.427 ± 0.380 | 1.00037 ± 0.00007 |

## 2. Experimental setup and imaging system

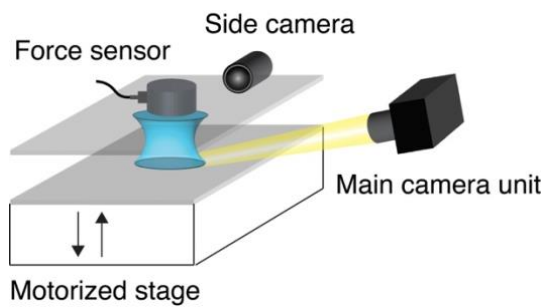

**Figure S1.** Schematics of the experimental setup and imaging system. A capillary bridge is formed between two parallel plates. The bottom plate is mounted on a motorized z-stage, which moves to compress or stretch the bridge at set speeds and distances, while the top plate is fixed and equipped with a force sensor to record the force on the upper surface. The system is imaged using a dual camera setup: a main high magnification camera unit (digital camera with zoom and magnifying lens) that focuses on the bridge edges to extract geometrical parameters (contact angles, contact radii, and bridge profile). Here, the focus is set on the bottom edge of the bridge as an example; during measurements, the top edge is similarly imaged when extracting the associated parameters (see SI 4 for details). A secondary side view camera provides a complementary full profile view of the bridge. The z-stage, force sensor, and both cameras operate in synchrony. Further details of the apparatus can be found in Goodband et al.[2]

## 3. Droplet placement for capillary bridge formation

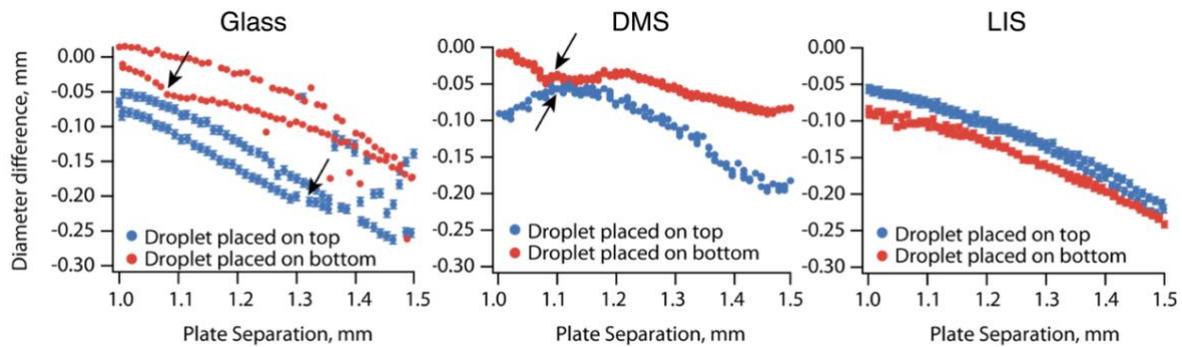

**Figure S2**. Variation of the difference between the top and bottom contact diameters of the capillary bridge over an extension-compression cycle. In all cases, a 10 µL droplet of glycerol solution is initially placed either on the top surface (blue) or bottom surface (red) before forming a capillary bridge with the other surface. The difference is negative for most surface separations, indicating a larger bottom diameter. This consistent with the expected effect of gravity. However, on Glass, the difference can initially be positive at small separation, depending on where the droplet is initially placed. Additionally, for Glass and DMS, the largest different is observed when the droplet is initially placed on the top surface. These observations are counterintuitive considering how gravity breaks the capillary bridge symmetry and point to pinning effects. On both Glass and DMS, significant pinning events can be seen during the experiment (see arrows). On LIS, the system behaves consistently as expected when considering gravity and in the absence of pinning. The droplet position at start does not affect the final capillary bridge behavior.

## 4. Focusing on the top and bottom of capillary bridges

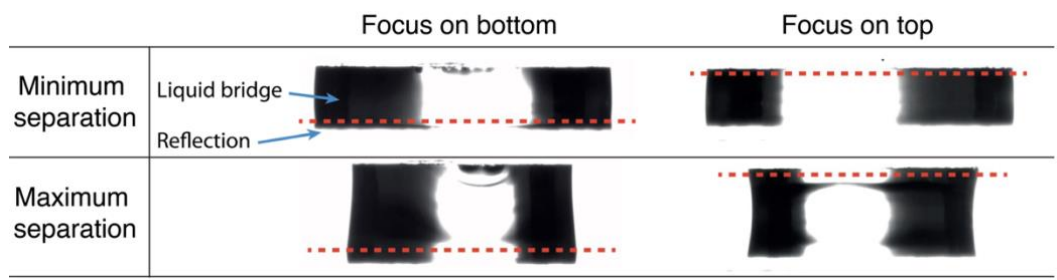

**Figure S3**. Example of sequential measurements being taken at the bottom (left) and top (right) surfaces. In order get a better view of the capillary bridge–surface contact region, the camera is placed at an angle.[2] This makes it is challenging to simultaneously and accurately measure the top and bottom of the bridge (assuming a 1 camera setup). For experiments that require information from both extremities, the measurements are conducted sequentially as illustrated here: multiple extension-compression cycles are obtained focusing on the bottom of the bridge (left) and subsequently on the top interface (right) of the same capillary bridge (see the next Supporting Information SI 5 for confirmation of the data reproducibility). The above example is taken on DMS and the bridge can be seen to have a reflection on either the top or bottom surface. A red line denotes the contact line of the bridge with the surface. This allows for an accurate detection of the bridge intersection with the surface, provided it is in focus.

# 5. Reproducibility of extension-compression cycles

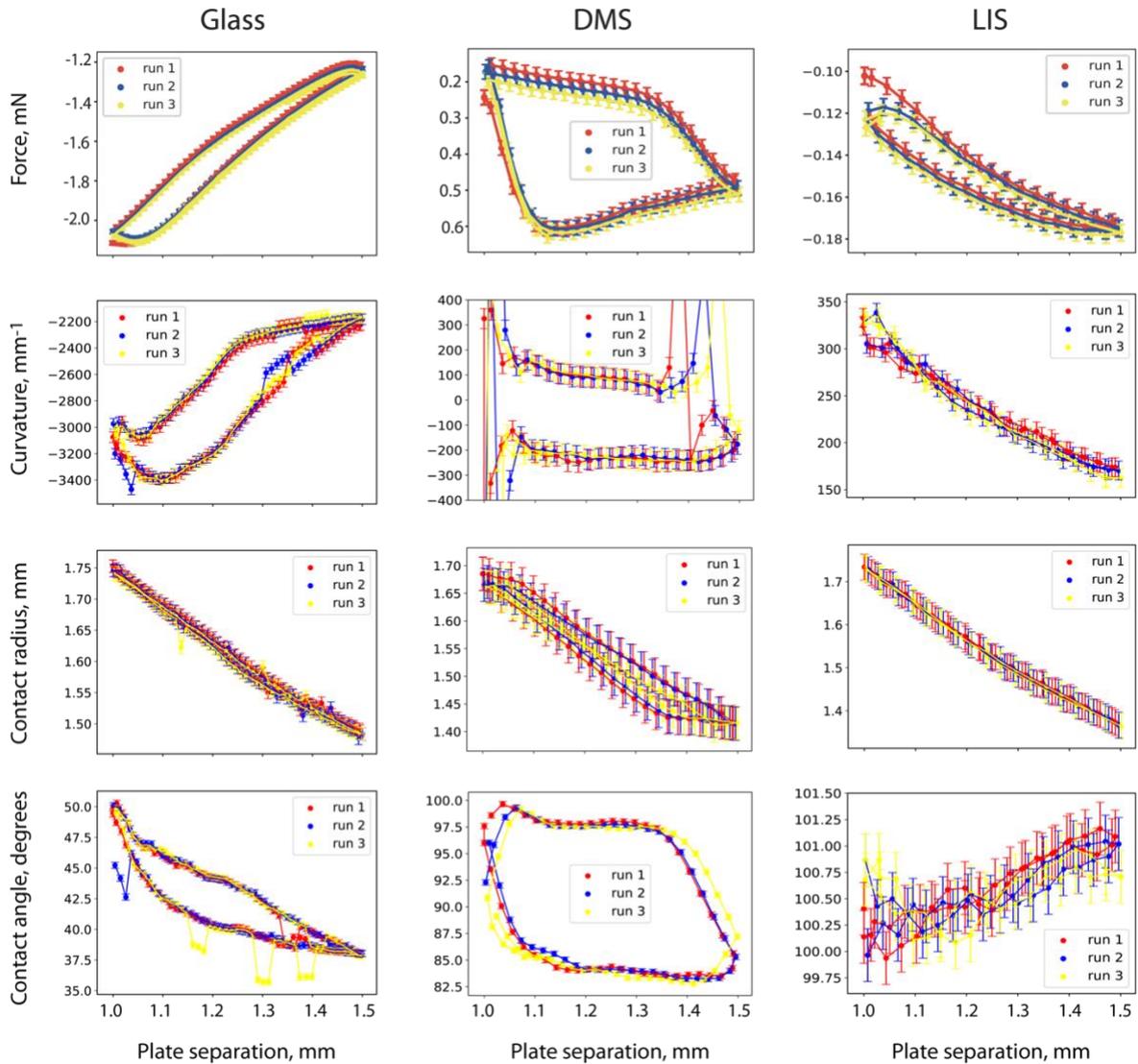

**Figure S4**. Evolution of the force, curvature, contact radius and contact angle for each system over a number of extension-compression cycles. In all cases the camera is focusing on the top part of the bridge. For each set of surfaces, the measurements are fully reproducible within error between consecutive cycles. This confirms the suitability of a sequential strategy for investigating the behaviour of the bottom and the top of a particular capillary bridge.

## 6. Acquisition of the radii of curvature

The radii of curvature are obtained at the plate (top or bottom) where the capillary force is being calculated. Two components are measured: the meridional radius of curvature, $R_2$, and the azimuthal radius of curvature, $R_1$. The detailed method for processing the experimental data can be found in Goodband et al.[2], and a similar calculation was employed by Wang et al.[3] Here, we briefly summarize the procedure as follows. To obtain the meridional radius of curvature ($R_2$), the local edge of the capillary bridge is fitted to a second-order polynomial. From this fit, the meridional curvature is computed as:[4]

$$K_2 = \frac{\left|\frac{d^2 y}{dx^2}\right|}{(1+(\frac{dy}{dx})^2)^{3/2}}$$

and the corresponding meridional radius of curvature is simply $R_2 = \frac{1}{K_2}$.

The azimuthal radius of curvature ($R_1$) is obtained from geometric relations at the three-phase contact line. At the top and bottom plates, respectively:

$$\frac{1}{R_1} = \frac{\sin\theta_t}{R_t} \quad \text{and} \quad \frac{1}{R_1} = \frac{\sin\theta_b}{R_b}.$$

Here, $R_t$ and $R_b$ are the top and bottom contact radii, and $\theta_t$ and $\theta_b$ are the corresponding contact angles as shown in Fig. 1 of the main text.

## 7. Impact of gravity

The capillary bridge used in this study are several millimeters wide for a height varying between 1 and 1.4 mm. This is relatively close to the capillary length of the bridge's solution (2.39 mm). Alternatively, we find the bond number $B_o$ of the capillary bridge at $0.1 < B_o < 0.4$, relatively close to 1. Gravitational effects must therefore be taken into account not only in terms of additional weight on the bottom surface, but also for its deformation of the bridge geometry.

**Bond number calculation**

The Bond number ($B_o$) is defined as the ratio of gravitational to surface tension forces with a Bo value of less than one indicating that surface tension dominates over gravity. It is given by the following equation:

$$B_o = \frac{gL^2(\rho_L - \rho_g)}{\sigma}$$

where $g$ is the acceleration due to gravity (9.81 ms$^{-2}$), $L$ is the characteristic length of the system (here taken as the bridge height[5]), $\rho_L$ is the liquid density, $\rho_g$ is the gas density, and $\sigma$ is the surface tension of the liquid. For the present study we find $B_o = 0.175$ for a capillary bridge at minimum extension (1 mm), and $B_o = 0.394$ for a capillary bridge at maximum extension (1.5 mm). While both $B_o$ numbers are smaller than 1, they indicate that gravitational effects still play a role that should be carefully considered in this setup.

## 8. Oil induced pinning on DMS

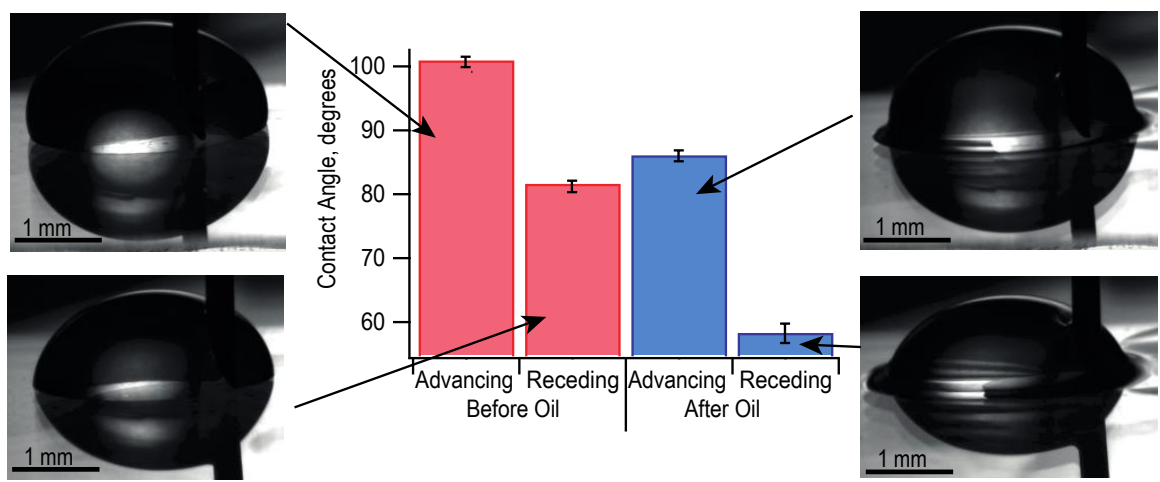

**Figure S5**. Silicone oil-induced droplet pinning on DMS. Representative advancing (top) and receding (bottom) contact angles are shown before (left) and after (right) the lubricant (silicone oil) contacted the base of a droplet of glycerol solution. As the oil contacts the base of the droplet, it spontaneously cloaks it (see Supplementary video S1) resulting in a reduction of both the advancing and receding angles. The decrease is however more marked for the receding angle (histograms) due to pinning of the contact line by the oil ridge. The ridge forces the droplet to reduce its CA with the DMS as liquid is being pumped out of the droplet. The whole experiment can be visualised as a video (Supplementary video S1). The analysis software developed for extracting geometrical parameters from the capillary bridges[2] cannot be applied here and the CA measurements were done manually using the angle tool in ImageJ[6]. Comparative testing on sample capillary bridge images typically evidenced a 2 degrees discrepancy between these two methods, with the capillary bridge fitting method in being more accurate on capillary bridges[2]. This effect, together with the experimental variability (see Methods section) explain the difference in CA between this figure and Figs. 5 and 6 of the main text. Each set of data is however consistent within itself, with the error bars shown representing two standard errors.

**Video S1**: Behaviour evolution of a droplet of glycerol solution placed on a DMS surface as silicone oil (the lubricant) diffuses towards the droplet base. In order to illustrate the advancing and receding CAs of the droplet with the DMS, 50 µL of the glycerol solution is constantly pumped in and out of the droplet using a motorised syringe pump. The pumping in (out) of the glycerol solution follows a linear ramp. A 40 µL drop of silicon oil is deposited on the DMS surface, several millimetres away from the drop of glycerol solution and allowed to diffuse in all directions. As the oil contacts the drop, it immediately and spontaneously cloaks it resulting in a change of the apparent advancing and receding CAs.

## 9. Contact line displacement over an extension-compression cycle

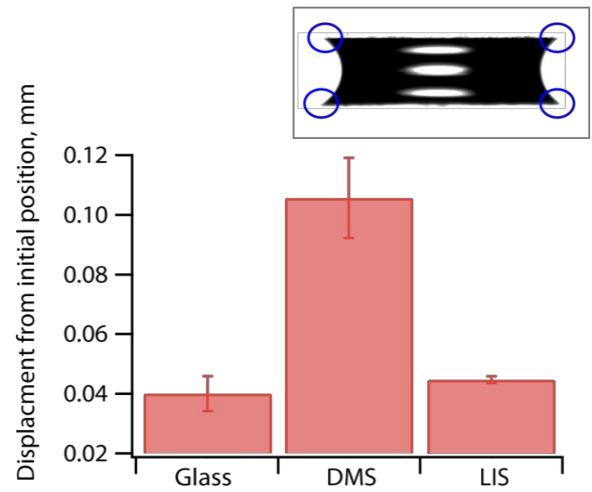

**Figure S6**. Variation of contact line position between the beginning and the end of a complete extension-compression cycle. The data represents the average displacement of the contact line over 4 separate points (blue highlights in the inset image). On Glass, very little displacement occurs due to high pinning (as shown in Fig. 3g of the main text and Fig. S4 of SI 5), the contact angle constantly changes. The measurement uncertainty represents the uneven movement of the contact lines, where one side may preferentially move more than the other. On DMS, the capillary bridge experiences pinning and a large displacement (sliding). Uneven movement caused by pinning leads to greater displacement and larger errors. In contrast, on LIS, the capillary bridge slides very slowly during the course of the measurement, with all contact lines moving at the same rate, hence resulting in a small error.

## 10. Effective surface tension for a cloaked capillary bridge

Deriving an effective surface tension for a droplet or liquid bridge which is cloaked is not straightforward because the effective surface tension of a cloaked liquid is known to change with the thickness of the cloaking thin film[7]. In our system, the lubricant film on the bridge cannot be measured directly. Furthermore, the film thickness may be non-uniform and can evolve dynamically, including through lubricant transport between the two surfaces via the capillary bridge. To estimate an effective surface tension, we therefore employ the cloaking film approximation:[8]

$$\gamma_c = \gamma_{dl} + \gamma_{la}$$

where $\gamma_c$ is the effective (cloaked) surface tension, $\gamma_{dl}$ is the droplet–lubricant interfacial tension, and $\gamma_{la}$ is the lubricant–air surface tension. For our system, $\gamma_{dl} = 27.6\ mN/m$ for an 80 wt% glycerol–water droplet in contact with 20 cSt silicone oil, and $\gamma_{la} = 20.6\ mN/m$ for 20 cst silicone oil in air, both measured using the pendant drop method[9,10]. This yiedls $\gamma_c = 48.2\ mN/m$. An approximate sum from literature value[8] for a similar system gives $\gamma_c = 52.9\ mN/m$. Averaging these two estimates, we adopt an effective droplet (cloaked)–gas surface tension of

$$\gamma_{dg} \sim 50\ mN/m.$$

## 11. Fitting capillary liquid bridges with oil ridges

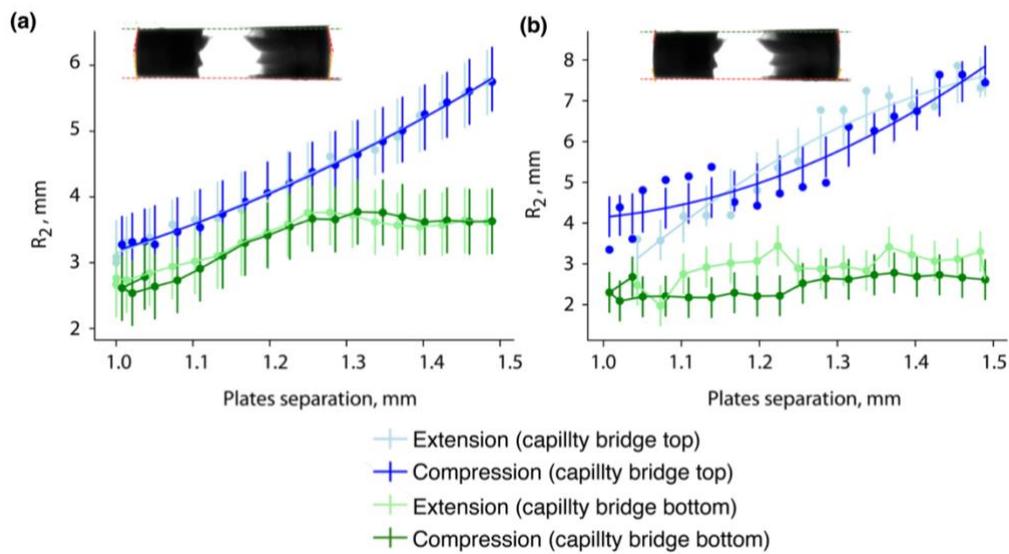

**Figure S7**. Impact of the selected region of the capillry bridge profile used in the fitting[2] carried out to derive the bridge's meridional radius of curvature $R_2$. To illustrate the issue, the upper graph (a) shows the curvature derived from fitting half of the bridge's height, whereas only a quarter of the bridge's height is taken in the lower graph (b). In each graph, the curvature is shown for both the top surface (blue) and bottom surface (green) with the extension data shown darker than the retraction in order to aid visualisation. The example is taken for a capillry bridge between two LIS surfaces. The error bars on all curves represent 0.5 mm.

## 12. Capillary bridge stick–slip motion with contact line pinning

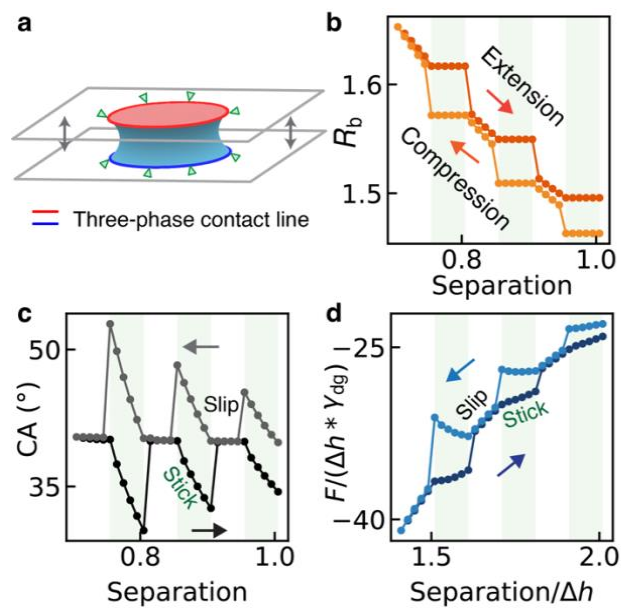

**Figure S8.** Simulation of the stick–slip contact line motion on a non-ideal surface. Here, this is achieved by employing three-phase contact line pinning and release during a capillary bridge compression-extension cycle, as illustrated in (a). The green shaded area indicates where the contact line is pinned, and white area indicates where the contact line is free to move. The change in the bridge bottom radius $R_b$ is shown in (b), the measured contact angle close to the bottom contact line is shown in (c), and the normalized measured capillary force through simulation is shown in (d).

To investigate the generality of the stick–slip behavior, we explored an alternative scenario of contact line pinning and depinning. We introduced pinning and release of the three-phase contact line on both plates simultaneously in an extension-compression cycle (Fig. S8a). When the contact line is free to move, the bottom contact radius $R_b$ increases during compression and decreases during extension (Fig. S8b), with the measured CA near the contact line remaining constant and close to the input parameters (Fig. S8c). When the contact line is pinned, the bottom radius remains unchanged (Fig. S8b), and the CA increases during compression and decreases during extension as the plates move (Fig. S8c). The change in CA is more pronounced at smaller plate separation, where the bridge is more compressed and constrained such that pinning triggers a stronger response. The capillary force in the simulation was calculated by Eq. 1 with the pressure obtained from the simulation model and normalized by the product of the plate separation change $\Delta h$ and the droplet–gas interfacial tension $\gamma_{dg}$. The resulting force (Fig. S8d) exhibits a stepwise behavior similar to that visible for Glass in Fig. 3j and that simulated in Fig. 4e for chemically heterogeneous surfaces.

S-13

## 13. Lubricant transfer from LIS to DMS in an asymmetric capillary bridge system

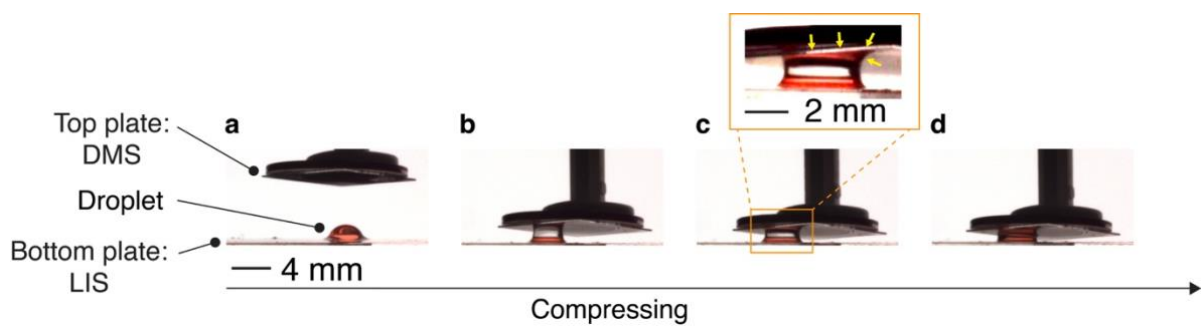

**Figure S9**. Formation and compression of a capillary bridge between a dyed silicone-oil-infused LIS (bottom plate) and a DMS surface (top plate). The lubricant infused into the LIS is 20 cSt silicone oil dyed with Oil Red O; this dyed lubricant visibly cloaks the deposited droplet (10 μL, 80 wt% glycerol in water), as shown in (a). The bottom plate is then raised to form a capillary bridge, with a dyed oil ridge present at the LIS–bridge contact. Upon compression (b–d), an oil ridge forms at the bridge contact with the top DMS surface, highlighted by yellow arrows in the inset of (c). Note that an oil ridge is also present in (b), though it may not be clearly visible due to the viewing angle. At smaller plate separations (d), the entire bridge becomes cloaked by the dyed lubricant, indicating that the lubricant has reached and interacted with the top DMS surface.

To directly examine whether lubricant transfers from the LIS to the DMS surface in the asymmetric capillary bridge configuration, we conduct a visualisation experiment using a dyed silicone oil infused into the LIS. The goal is to qualitatively assess lubricant migration independent of gravity-driven effects. The dyed lubricant is prepared by dissolving 1 wt% Oil Red O in 20 cSt silicone oil, assisted by sonication and filtered through a 0.2 μm syringe filter to remove undissolved dye. This lubricant is then infused into the LIS. Owing to the limited sensitivity of the digital camera, the LIS is loaded with ~50% more dyed lubricant than the standard protocol to ensure that the oil ridge was optically detectable.

A droplet is deposited onto the dyed LIS, where it becomes visibly cloaked by the lubricant. The bottom (LIS) plate is then brought into gentle contact with the top DMS substrate to form a capillary bridge, during which a dyed oil ridge is clearly visible on the LIS side. When the bridge was compressed to heights comparable to those studied in the main text (~1.72 mm in Fig. S9c and ~0.94 mm in Fig. S9d; the relevant range in the main text is ~1–1.5 mm), an oil ridge also form at the bridge–DMS contact (Fig. S9c, highlighted with yellow arrows in the inset). At smaller separations, the entire bridge gradually turned red (Fig. S9d), indicating that the dyed lubricant cloaked the full bridge interface.

This experiment is qualitative: the film thickness and detailed lubricant flow could not be resolved with the current optical setup, and the manual operation involved higher velocities (≫ 0.008 mm s$^{-1}$ used elsewhere in this work), and increased lubricant loading compared to the main experiments. Nevertheless, these observations support the interpretation that lubricant exchange occurs during asymmetric capillary bridge deformation. Quantitative characterisation will require higher resolution techniques such as confocal microscopy with fluorescent dyes, something beyond the scope of the current work.

## 14. Simulation and experiment comparison for LIS-top-DMS-bottom capillary bridge

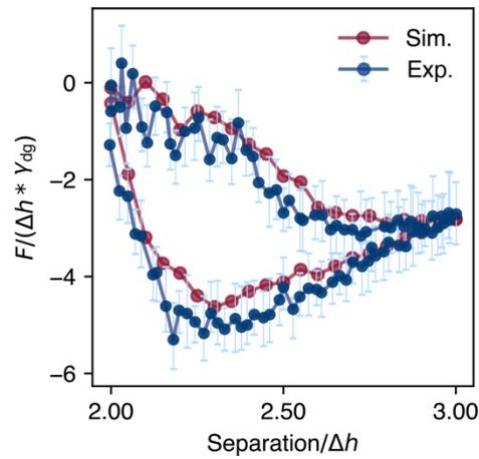

**Figure S10**. Normalized force versus separation for a capillary bridge with the LIS at top and DMS at bottom. The simulated force is obtained from the capillary bridge pressure output following Eq. 1 of the main text, while the experimental force is inferred at the bottom surface using Eq. 4 of the main text. The good agreement between experiments and simulations indicates that the model accurately captures the behavior of asymmetric capillary bridges.

## 15. References

(1) Nečas, D.; Klapetek, P. Gwyddion: An Open-Source Software for SPM Data Analysis. *Cent. Eur. J. Phys.* **2012**, *10* (1), 181–188.

(2) Goodband, S.; Kusumaatmaja, H.; Voïtchovsky, K. Development of a Setup to Characterize Capillary Liquid Bridges between Liquid Infused Surfaces. *AIP Adv.* **2022**, *12*, 015120.

(3) Wang, Q.; Chen, W.; Wu, J. Effect of Capillary Bridges on the Interfacial Adhesion of Wearable Electronics to Epidermis. *Int. J. Solids Struct.* **2019**, *174–175*, 85–97.

(4) do Carmo, M. P. *Differential Geometry of Curves and Surfaces: Revised and Updated Second Edition*; Courier Dover Publications, 2016.

(5) Radoev, B.; Petkov, P.; Ivanov, I. Capillary Bridges — A Tool for Three-Phase Contact Investigation; IntechOpen, 2015; p 32.

(6) Schneider, C. A.; Rasband, W. S.; Eliceiri, K. W. NIH Image to ImageJ: 25 Years of Image Analysis. *Nat. Methods* **2012**, *9* (7), 671–675.

(7) Pepper, K. G.; Bahrim, C.; Tadmor, R. Interfacial Tension and Spreading Coefficient of Thin Films: Review and Future Directions. *J. Adhes. Sci. Technol.* **2011**, *25* (12), 1379–1391.

(8) Gunjan, M. R.; Kumar, A.; Raj, R. Cloaked Droplets on Lubricant-Infused Surfaces: Union of Constant Mean Curvature Interfaces Dictated by Thin-Film Tension. *Langmuir* **2021**, *37* (22), 6601–6612.

(9) Berry, J. D.; Neeson, M. J.; Dagastine, R. R.; Chan, D. Y. C.; Tabor, R. F. Measurement of Surface and Interfacial Tension Using Pendant Drop Tensiometry. *J. Colloid Interface Sci.* **2015**, *454*, 226–237.

(10) Daerr, A.; Mogne, A. Pendent_Drop: An ImageJ Plugin to Measure the Surface Tension from an Image of a Pendent Drop. *J. Open Res. Softw.* **2016**, *4* (1), e3.